\documentclass[preprintnumbers,amsmath,amssymb]{revtex4-1}

\usepackage{epsfig}
\usepackage{graphicx}
\usepackage{subfigure}
\usepackage{float}
\usepackage{multirow}
\usepackage{amssymb}
\usepackage{amsmath}
\usepackage{picinpar}

\begin{document}


\newcommand{\figwidth}{1.0\columnwidth}
\newcommand{\subfigwidth}{0.7\columnwidth}
\newcommand{\subfigwidthtwo}{0.45\columnwidth}


\newcommand{\smin}{{\mbox{\scriptsize{min}}}}
\newcommand{\smax}{{\mbox{\scriptsize{max}}}}
\newcommand{\stot}{{\mbox{\scriptsize{tot}}}}
\newcommand{\tot}{{\mbox{\scriptsize{tot}}}}

\newcommand{\ms[2]}{{\mbox{\scriptsize{#2}}}}
\newcommand{\im}{i}
\newcommand{\jm}{j}

\newcommand{\mA}{\langle m_A\rangle}
\newcommand{\mB}{{\langle m_B\rangle}}
\newcommand{\pmB}{{\langle pm_B\rangle}}
\newcommand{\m}{{\langle m_\tot\rangle}}

\title{Compensation temperature in spin-$1/2$ Ising trilayers: A Monte Carlo study}

\author{I. J. L. Diaz}\email{ianlopezdiaz@gmail.com}
\affiliation{Departamento de F\'{i}sica,
Universidade Federal de Santa Catarina,
88040-900, Florian\'{o}polis, SC, Brazil}
\author{N. S. Branco}\email{nsbranco@gmail.com}
\affiliation{Departamento de F\'{i}sica,
Universidade Federal de Santa Catarina,
88040-900, Florian\'{o}polis, SC, Brazil}

\date{\today}

\begin{abstract}
We study the magnetic and thermodynamic properties of a spin-$1/2$ Ising system
containing three layers, each of which is composed exclusively of
one out of two possible types of atoms, \textbf{A} or \textbf{B}.
The \textbf{A-A} and \textbf{B-B} bonds are ferromagnetic
while the \textbf{A-B} bonds are antiferromagnetic.
The study is performed through Monte Carlo simulations using the Wolff algorithm
and the data are analyzed with the aid of the multiple-histogram reweighting technique and finite-size scaling tools.
We verify the occurrence of a compensation phenomenon and obtain the compensation and critical
temperatures of the model as functions of the Hamiltonian parameters.
The influence of each parameter on the overall behavior of the system is discussed in detail
and we present our results in the form of phase diagrams dividing the parameter space in regions
where the compensation phenomenon is present or absent.
Our results may provide invaluable information for experimentalists seeking to build
materials with desired characteristics.
\end{abstract}

\pacs{05.10.Ln; 05.50.+q; 75.10.Hk; 75.50.Gg}

\maketitle

\section{Introduction}\label{introduction}

The industry's commitment to producing ever-smaller electronic devices
has contributed significantly to the increase in interest in the study of properties
of thin films and other nanoscopic materials.
From both theoretical and experimental points of view,
there are many interesting behaviors and unusual phase diagrams that
may arise when a magnet is composed of multiple layers with different magnetic properties.
For instance, the existence of antiferromagnetic couplings
between adjacent ferromagnetic layers generates a number of effects
with great potential for important technological applications such as in
magneto-optical recordings \cite{connell1982magneto},
spintronics \cite{grumberg1986layered},
the giant magnetoresistance \cite{camley1989theory},
and the magnetocaloric effects \cite{phan2007review}.
A particularly interesting phenomenon related to this type of layered ferrimagnets 
is the existence of compensation points, i.e., temperatures below the critical point for which the total magnetization is zero
while the individual layers remain magnetically ordered \cite{cullity2011introduction}.
The fact that the compensation point of some ferrimagnets occurs near room temperature
makes them particularly important for applications in magneto-optical drives \cite{connell1982magneto}.
Interestingly, certain physical properties, such as coercivity,
may exhibit a singular behavior at the compensation point,
even though the compensation phenomenon is completely unrelated to criticality
\cite{connell1982magneto, shieh1986magneto, ostorero1994dy}.


The development and improvement of thin film growth techniques, namely
molecular-beam epitaxy (MBE) \cite{herman2012molecular},
metalorganic chemical vapor deposition (MOCVD) \cite{stringfellow1999organometallic},
pulsed laser deposition (PLD) \cite{singh1990pulsed, chrisey1994pulsed},
and atomic layer deposition (ALD) \cite{leskela2003atomic, george2010atomic},
have enabled the experimental realization of several layered materials with specific characteristics, such as
bilayer \cite{stier2011carrier},
trilayer \cite{smits2004antiferromagnetic, leiner2010observation},
and multilayer \cite{kepa2001antiferromagnetic,
chern2001antiparallel,
sankowski2005interlayer,
chung2011investigation,
samburskaya2013magnetization} systems.
Nevertheless, progress in experimental investigation of these materials
becomes a slow and difficult process without detailed theoretical studies to guide.
In this sense, several theoretical models (e. g,
spin-$1/2$ Ising,
spin-$1$ Ising,
mixed-spin Ising,
Potts,
and Heisenberg models)
have been used to advance the understanding of the properties of these magnetic systems.

Even the theoretical study of these systems has its difficulties,
considering that only a handful of them are exactly solvable \cite{baxter1982exactly};
we must therefore resort to approximation methods.
For instance, spin-$1/2$, spin-$1$ and mixed-spin Ising bilayers have been studied
within various approaches, such as
the mean-field approximation (MFA) \cite{lipowski1993layered, hansen1993two, kaneyoshi1995relation, oitimaa2005ferrimagnetism},
the effective-field approximation (EFA) \cite{kaneyoshi1993magnetic, javsvcur1995effect, ainane2007magnetic, bayram2008effective, bayram2011mixed, bayram2011phase, ersin2014magnetic},
renormalization group (RG) \cite{hansen1993two, li2001critical, mirza2003phenomenological},
transfer matrix (TM) \cite{lipowski1993layered, lipowski1998critical, li2001critical},
high-temperature series expansion \cite{oitimaa2005ferrimagnetism},
and Monte Carlo (MC) simulations \cite{hansen1993two, ferrenberg1991monte, ahmed2013monte, wang2016monte, wang2017compensation, diaz2017monte}.
A cellular automata (CA) simulation method was also employed in the analysis of both spin-$1/2$ Ising and $3$-states Potts model bilayers \cite{asgari2008obtaining}.
The Green's function (GF) method was used to study a Heisenberg bilayer \cite{ping2017magnetization},
and the pair approximation (PA) was also employed in the study
of spin-$1/2$ Heisenberg and Ising bilayers \cite{szalowski2013influence, balcerzak2014ferrimagnetism} and multilayers \cite{szalowski2012critical, szalowski2014normal}.
Monte Carlo simulations were also applied in the study of Heisenberg \cite{razouk2011monte} and Ising \cite{diaz2018monte} multilayer systems.

On the other hand, trilayers and three-layered superlattices have not yet been studied as extensively
as bilayers and multilayers.
Among the few examples found in the literature, we mention
an alternating spin-$1/2$, -$1$, and -$3/2$ Ising three-layered superlattice
which was studied both in a mean-field approach
\cite{naji2014phase} and through MC simulations
\cite{naji2014monte}.
We also cite a spin-$1/2$ Ising hexagonal lattice trilayer analyzed within the EFA
\cite{santos2017effective}
and a spin-$1/2$ Ising square lattice trilayer analyzed within both the MFA and the EFA \cite{diaz2018ferrimagnetism}.

In this work we are particularly interested in the type of system presented
in Refs. \onlinecite{santos2017effective,diaz2018ferrimagnetism},
since these works show, in an effective-field approach, that it is possible for a single-spin system with an odd number of layers
to exhibit a ferrimagnetic phase with compensation without site dilution,
as opposed to the case of single-spin bilayers \cite{balcerzak2014ferrimagnetism, diaz2017monte}
and multilayers \cite{szalowski2014normal, diaz2018monte}, for which dilution
is a necessary condition for the existence of a non-zero compensation temperature.
Therefore, we study a three-layer spin-$1/2$ Ising model with two types of atoms (\textbf{A}
and \textbf{B}, say), such that each layer is composed of only one type of atom.
Our goal is to establish the conditions for the appearance of the compensation effect
and the contribution of each parameter to the occurrence of said effect.
This model has already been analyzed in Ref. \onlinecite{diaz2018ferrimagnetism}
within the MFA and EFA approaches, however,
although mean-field-like approximations usually provide a fast and qualitatively right
solution for most models, they do not always describe the actual physical behavior of some low-dimensional systems
(see Ref. \onlinecite{boechat2002renormalization} and references therein).
Therefore, we conduct this study within a Monte Carlo approach, using the Wolff single-cluster algorithm \cite{artigo:wolff}
and with the aid of a reweighting multiple histogram technique \cite{artigo:ferrenberg:histograma1,artigo:ferrenberg:histograma2}
and finite-size scaling tools.
In Sec. \ref{sec:model} we present the model and discuss the simulation and data analysis methods.
We present our results and discussion in Sec. \ref{sec:results}.
In Sec. \ref{sec:conclusion}, the final remarks and conclusions are drawn.

\section{Model and Monte Carlo Simulations}\label{sec:model}

\begin{figure}[h]
\begin{center}
\subfigure[\textbf{AAB}\label{fig:01:a}]{
\includegraphics[width=\subfigwidth]{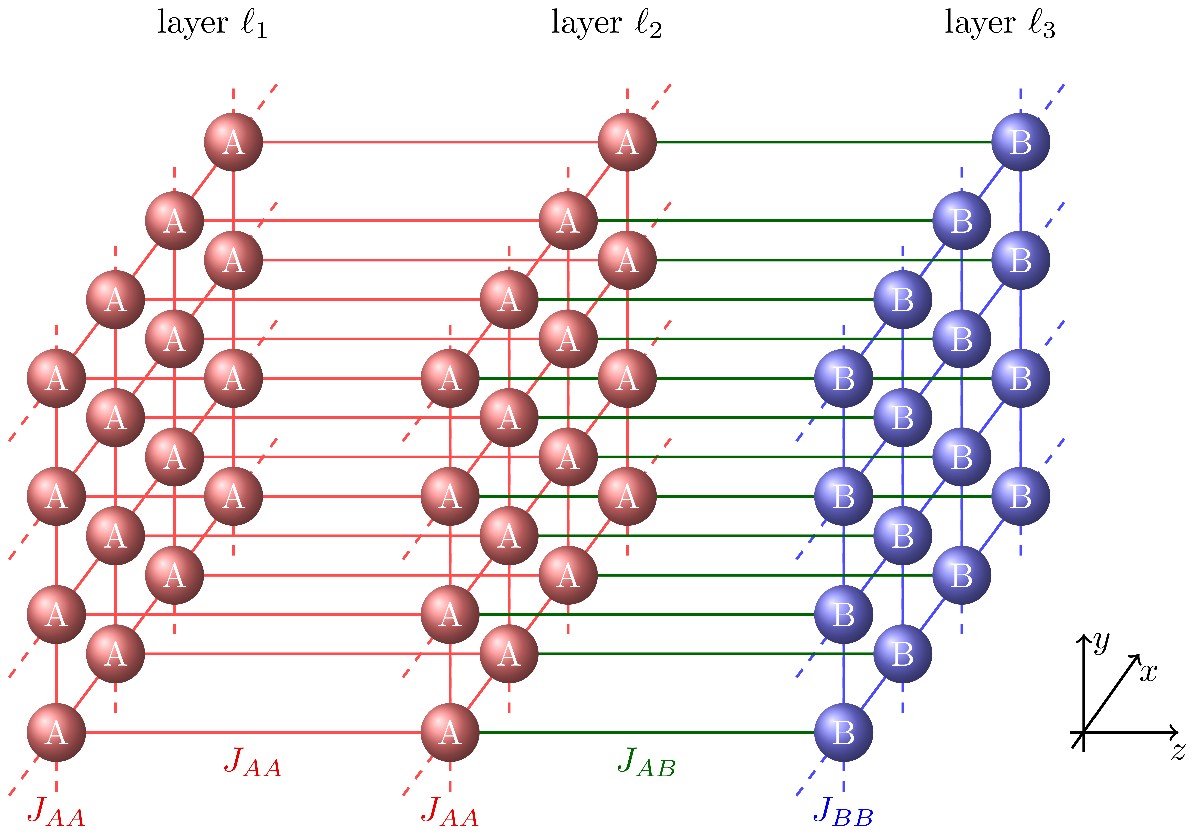}
}
\subfigure[\textbf{ABA}\label{fig:01:b}]{
\includegraphics[width=\subfigwidth]{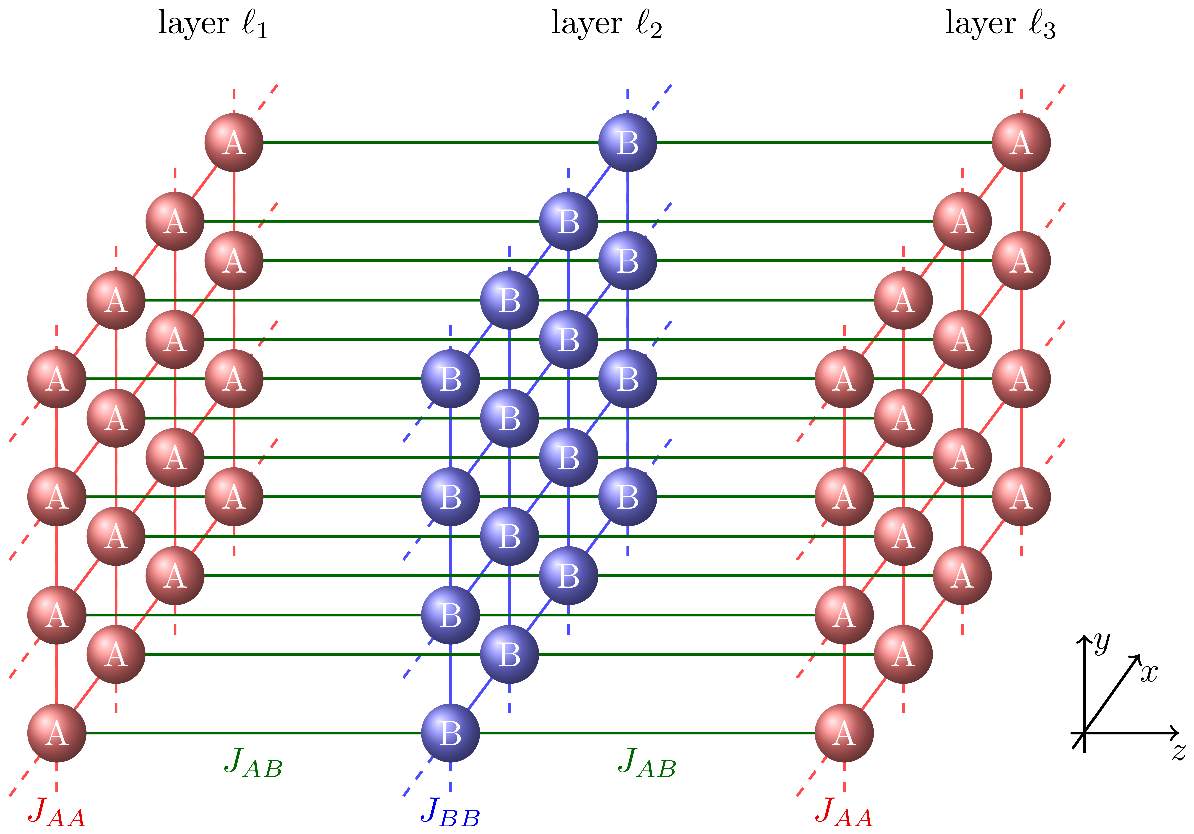}
}
\caption{
\label{fig:01}
A schematic representation of the trilayer systems.
In (a), we have the \textbf{AAB} system, in which
$J_{11}=J_{12}=J_{22}=J_{AA}>0$, $J_{23}=J_{AB}<0$, and $J_{33}=J_{BB}>0$.
In (b), we have the \textbf{ABA} system, in which $J_{11}=J_{33}=J_{AA}>0$; $J_{12}=J_{23}=J_{AB}<0$; $J_{22}=J_{BB}>0$.
}
\end{center}
\end{figure}

We study a trilayer system consisting of three monatomic layers, $\ell_1$, $\ell_2$, and $\ell_3$,
each of which is composed exclusively of either type-\textbf{A} or type-\textbf{B} atoms (see Fig. \ref{fig:01}).
The general system is described by the spin-1/2 Ising Hamiltonian
\begin{align}\label{eq:hamiltonian}
-\beta\mathcal{H}=
 K_{11}\sum_{\langle ii'\rangle}s_i s_{i'}
+K_{22}\sum_{\langle jj'\rangle}s_j s_{j'}
+K_{33}\sum_{\langle kk'\rangle}s_k s_{k'}
+K_{12}\sum_{\langle ij\rangle}s_i s_j
+K_{23}\sum_{\langle jk\rangle}s_j s_k,
\end{align}
where $\langle ii'\rangle$, $\langle jj'\rangle$, and $\langle kk'\rangle$
indicate summations over all pairs of nearest-neighbor sites in the same layer,
whereas $\langle ij\rangle$ and $\langle jk\rangle$
are over pairs of nearest-neighbor sites in adjacent layers.
The spin variables $s_n$ assume the values $\pm 1$,
the couplings are $K_{\delta\eta}\equiv \beta J_{\delta\eta}$,
where $\beta\equiv (k_BT)^{-1}$, $T$ is the temperature, $k_B$ is the Boltzmann constant,
and the exchange integrals $J_{\delta\eta}$ are $J_{AA}>0$ for \textbf{A-A} bonds,
$J_{BB}>0$ for \textbf{B-B} bonds, and $J_{AB}<0$ for \textbf{A-B} bonds.
We considered the two possible configurations of the trilayer
in which there are more atoms of type-\textbf{A} than type-\textbf{B} (see Fig. \ref{fig:01}),
namely, the \textbf{AAB} system is the case in which
$J_{11}=J_{12}=J_{22}=J_{AA}$, $J_{23}=J_{AB}$, and $J_{33}=J_{BB}$ (Fig. \ref{fig:01:a}),
whereas the \textbf{ABA} system corresponds to
$J_{11}=J_{33}=J_{AA}$, $J_{12}=J_{23}=J_{AB}$, and $J_{22}=J_{BB}$ (Fig. \ref{fig:01:b}).

For the Monte Carlo simulations, we employed the Wolff single-cluster algorithm \cite{artigo:wolff}
to analyze Hamiltonian \eqref{eq:hamiltonian} on a system of three stacked square lattices with $L^2$
sites each. We used periodic boundary conditions on the $xy$-plane and free boundary conditions in the $z$-direction (see Fig. \ref{fig:01}).
The Metropolis \cite{artigo:metropolis} dynamics was also implemented in early stages of the work,
but we preferred to use only the Wolff algorithm, considering that the single-spin flip dynamics
proved extremely inefficient for high \textbf{A-B} coupling asymmetry.
For both \textbf{AAB} and \textbf{ABA} trilayers,
we performed simulations for linear sizes $L$ from $10$ to $100$
and for a range of values of the Hamiltonian parameters:
$0.0<J_{AA}/J_{BB}\leq 1.0$, and $-1.0\leq J_{AB}/J_{BB}<0.0$.
The Mersenne Twister pseudo-random number generator \cite{artigo:mersenne-twister}
was used to generate all random numbers throughout the simulations.

At every step of the simulation (i.e. every single-cluster update),
we calculate the dimensionless energy $\mathcal{E}\equiv \mathcal{H}/J_{BB}$,
the magnetizations in each layer
\begin{equation}
\mathcal{M}_\lambda=\frac{1}{L^2}\sum_{n\in\ell_\lambda}s_n,
\end{equation}
where $\lambda=1,2,3$,
from which we obtain the total magnetization of the system, given by
\begin{equation}
\mathcal{M}_\tot=\frac{1}{3}(\mathcal{M}_1+\mathcal{M}_2+\mathcal{M}_3).
\end{equation}
The time series of $\mathcal{E}$ and $\mathcal{M}_\lambda$ were used to determine
the relevant integrated autocorrelation time $\tau$
after the initial $t_{eq}$ steps were discarded to account for thermalization \cite{livro:barkema}.
The number of steps performed in each simulation was always sufficient to generate
at least $10^4$ statistically independent states in order to calculate the canonical averages
$m_\Lambda\equiv\langle \mathcal{M}_\Lambda\rangle$, where $\Lambda=\tot,1,2,3$.
For instance, for the largest systems considered, i.e., $L=100$,
we performed up to $1.5\times 10^6$ steps and obtained $t_{eq}\approx 7.5\times 10^4$ and $\tau\approx 70$.
We also calculated the magnetic susceptibilities
\begin{equation}\label{eq:sus}
\chi_\Lambda=N_\Lambda K\left(\langle \mathcal{M}_\Lambda^2\rangle -\langle |\mathcal{M}_\Lambda|\rangle^2 \right),
\end{equation}
where $N_1=N_2=N_3=L^2$, $N_\tot=3L^2$, and $K\equiv J_{BB}(k_BT)^{-1}$ is the inverse dimensionless temperature.
The errors associated with the magnetizations and susceptibilities were
determined through the jackknife method \cite{livro:barkema}.

\begin{figure}[h]
\begin{center}
\subfigure[With compensation.\label{fig:mags:AAB:a}]{
\includegraphics[width=\subfigwidthtwo]{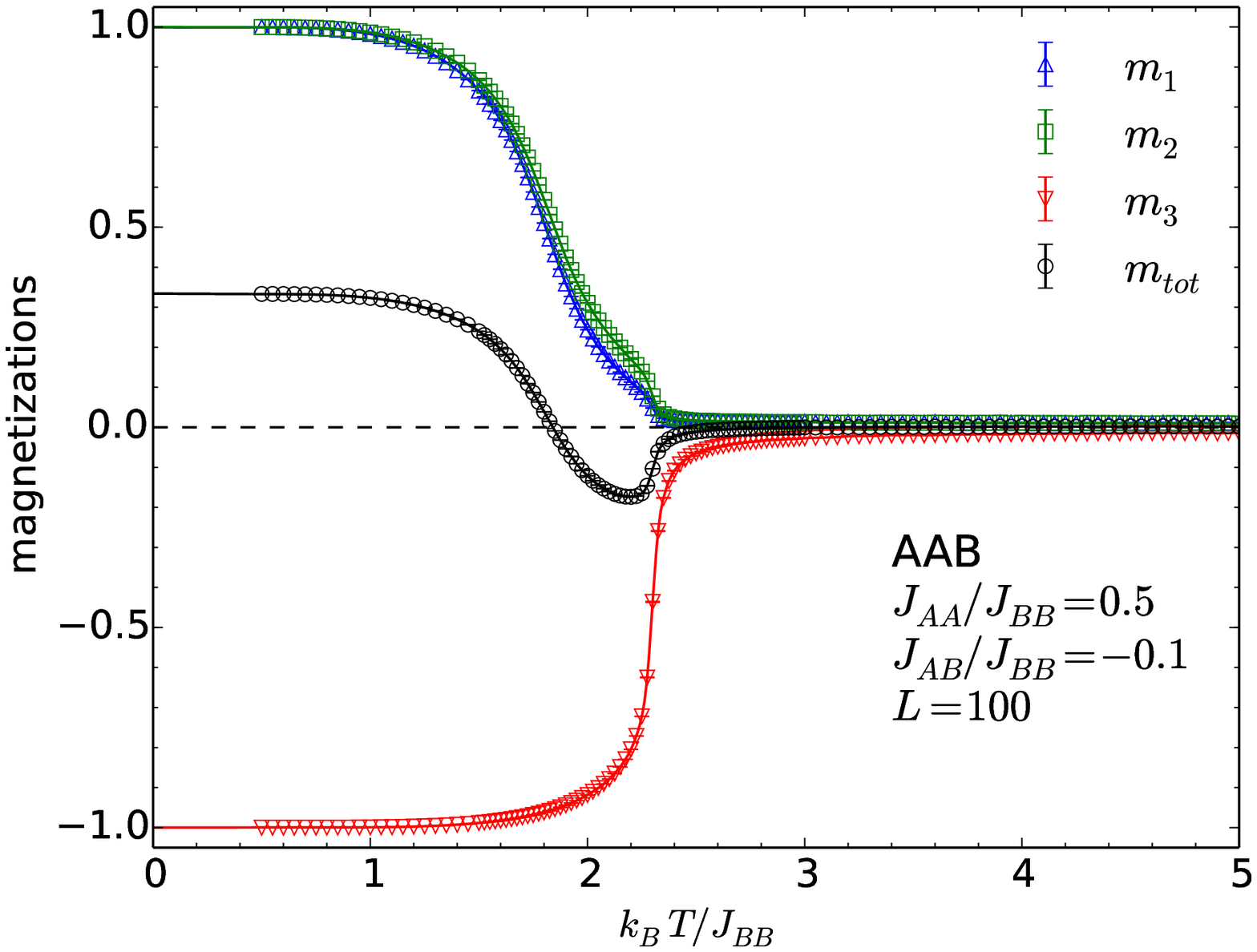}
}
\subfigure[Without compensation.\label{fig:mags:AAB:b}]{
\includegraphics[width=\subfigwidthtwo]{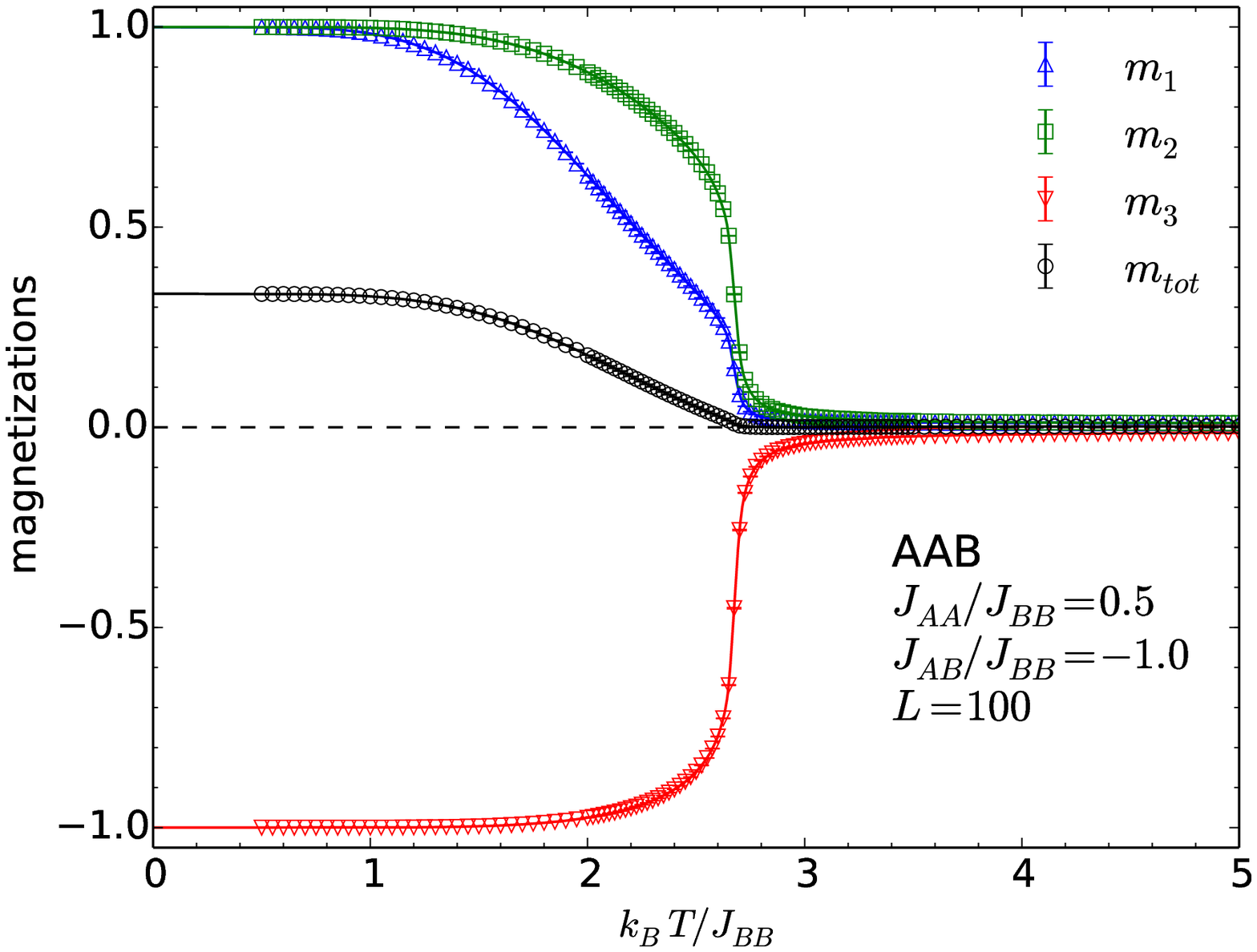}
}
\\
\subfigure[With compensation.\label{fig:mags:AAB:c}]{
\includegraphics[width=\subfigwidthtwo]{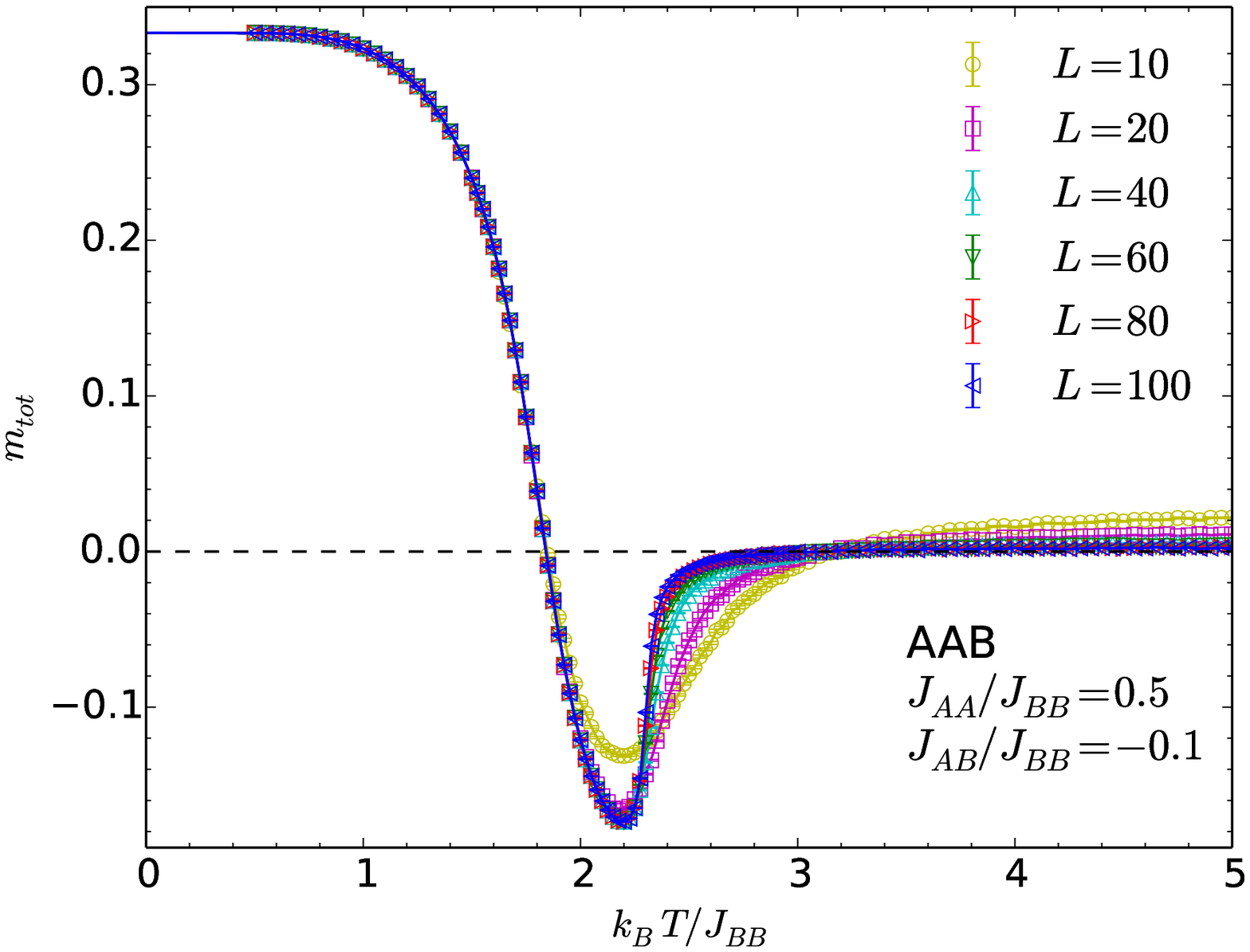}
}
\subfigure[Without compensation.\label{fig:mags:AAB:d}]{
\includegraphics[width=\subfigwidthtwo]{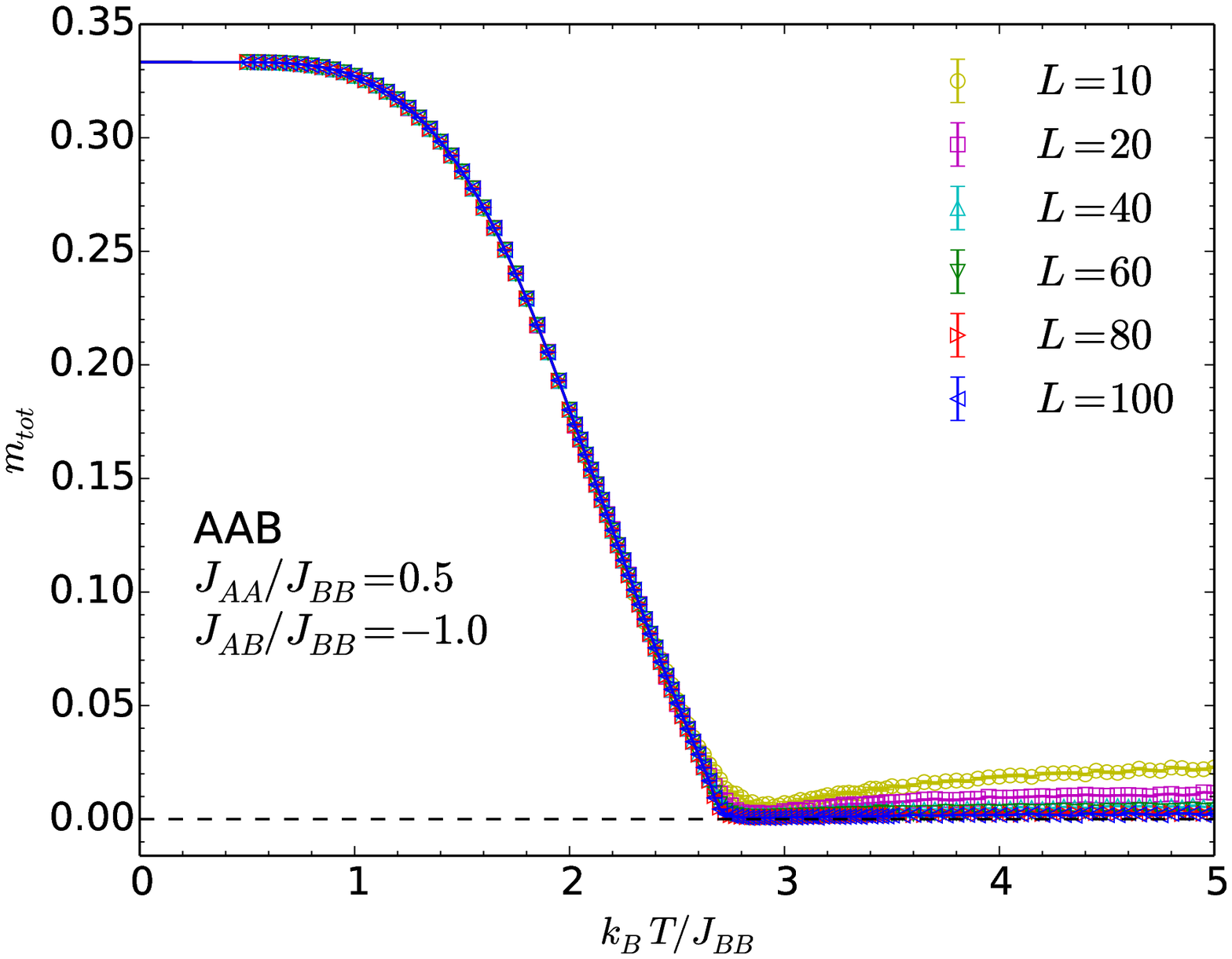}
}
\caption{
\label{fig:mags:AAB}
Magnetizations as functions of the dimensionless temperature $k_BT/J_{BB}$
for the \textbf{AAB} trilayer with $J_{AA}/J_{BB}=0.50$.
In (a) and (b) we show all magnetizations for $L=100$,
whereas for (c) and (d) we show only the total magnetization for several system sizes.
Figures (a) and (c), for $J_{AB}/J_{BB}=-0.1$,
show a compensation temperature $T_{comp}<T_c$ such that $m_\tot=0$.
Figures (b) and (d), for $J_{AB}/J_{BB}=-1.0$, show no compensation effect.
}
\end{center}
\end{figure}

\begin{figure}[h]
\begin{center}
\subfigure[With compensation.\label{fig:mags:ABA:a}]{
\includegraphics[width=\subfigwidthtwo]{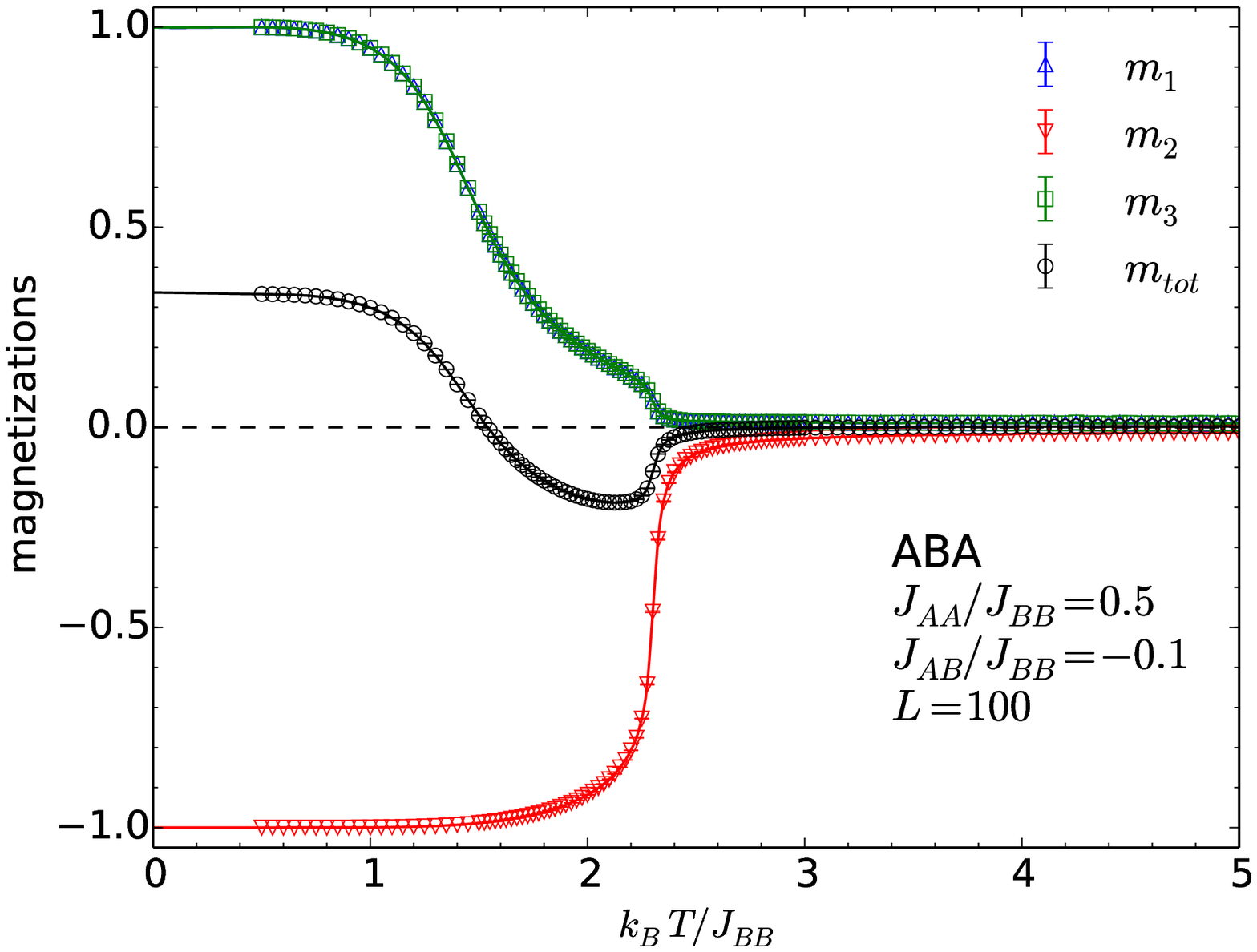}
}
\subfigure[Without compensation.\label{fig:mags:ABA:b}]{
\includegraphics[width=\subfigwidthtwo]{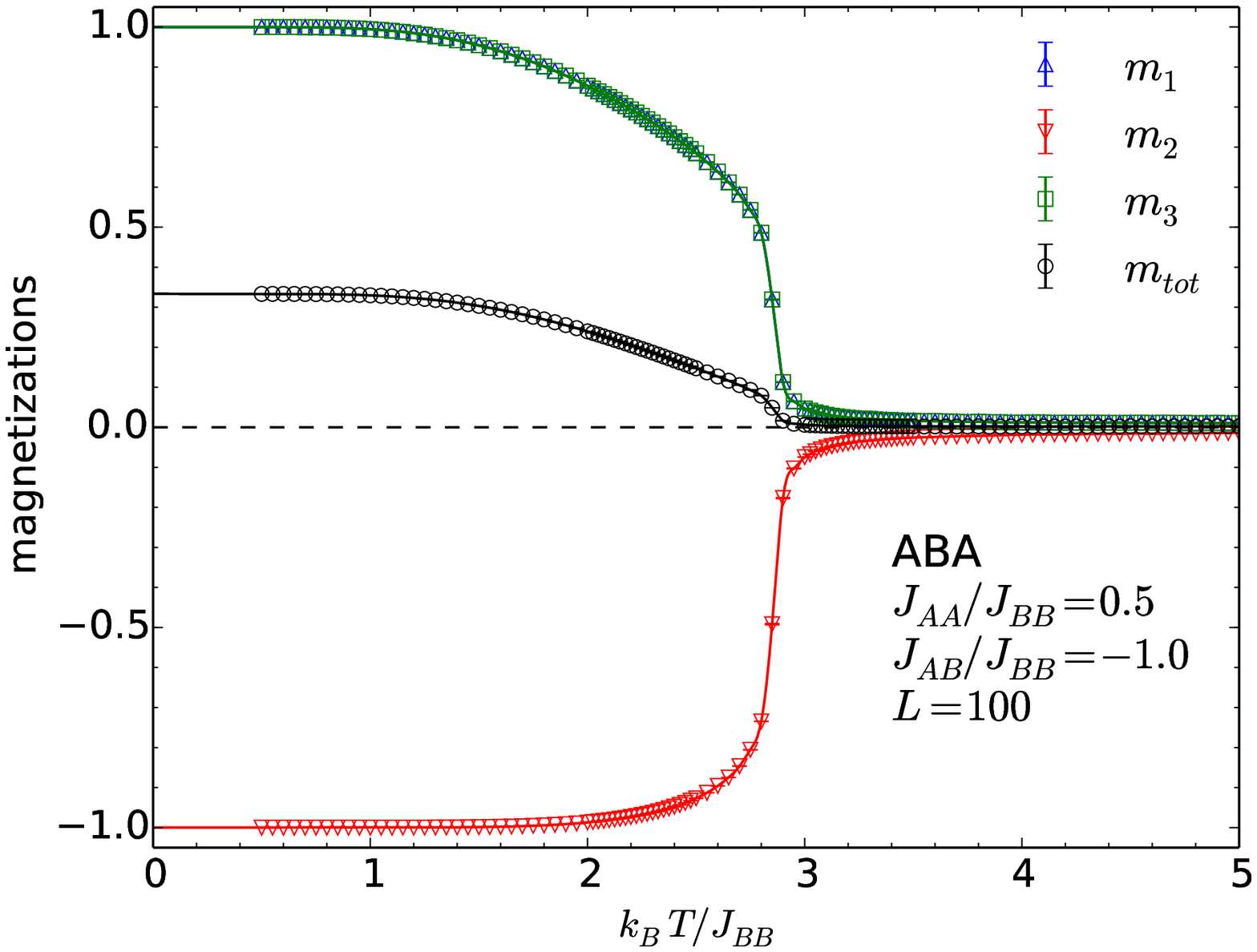}
}
\\
\subfigure[With compensation.\label{fig:mags:ABA:c}]{
\includegraphics[width=\subfigwidthtwo]{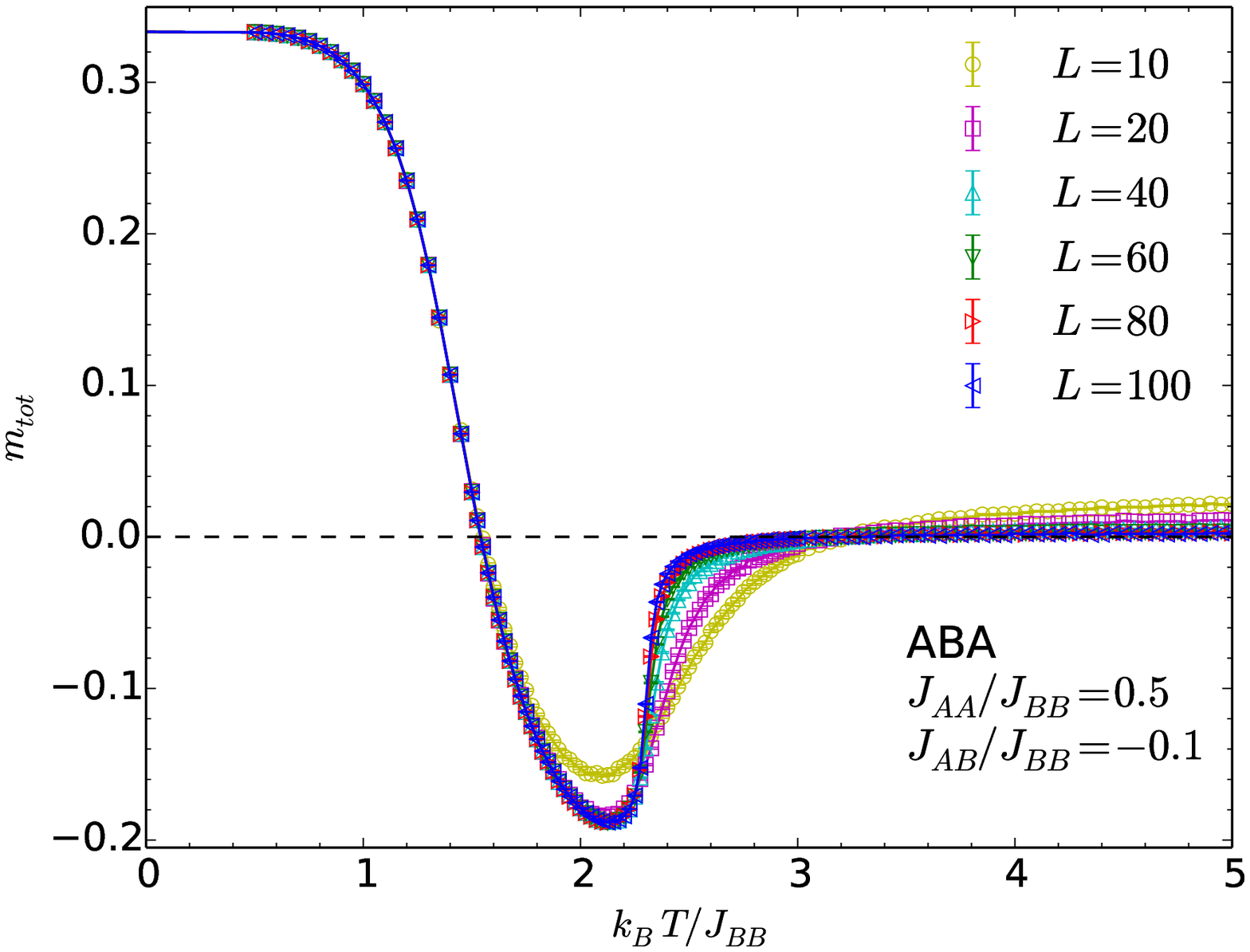}
}
\subfigure[Without compensation.\label{fig:mags:ABA:d}]{
\includegraphics[width=\subfigwidthtwo]{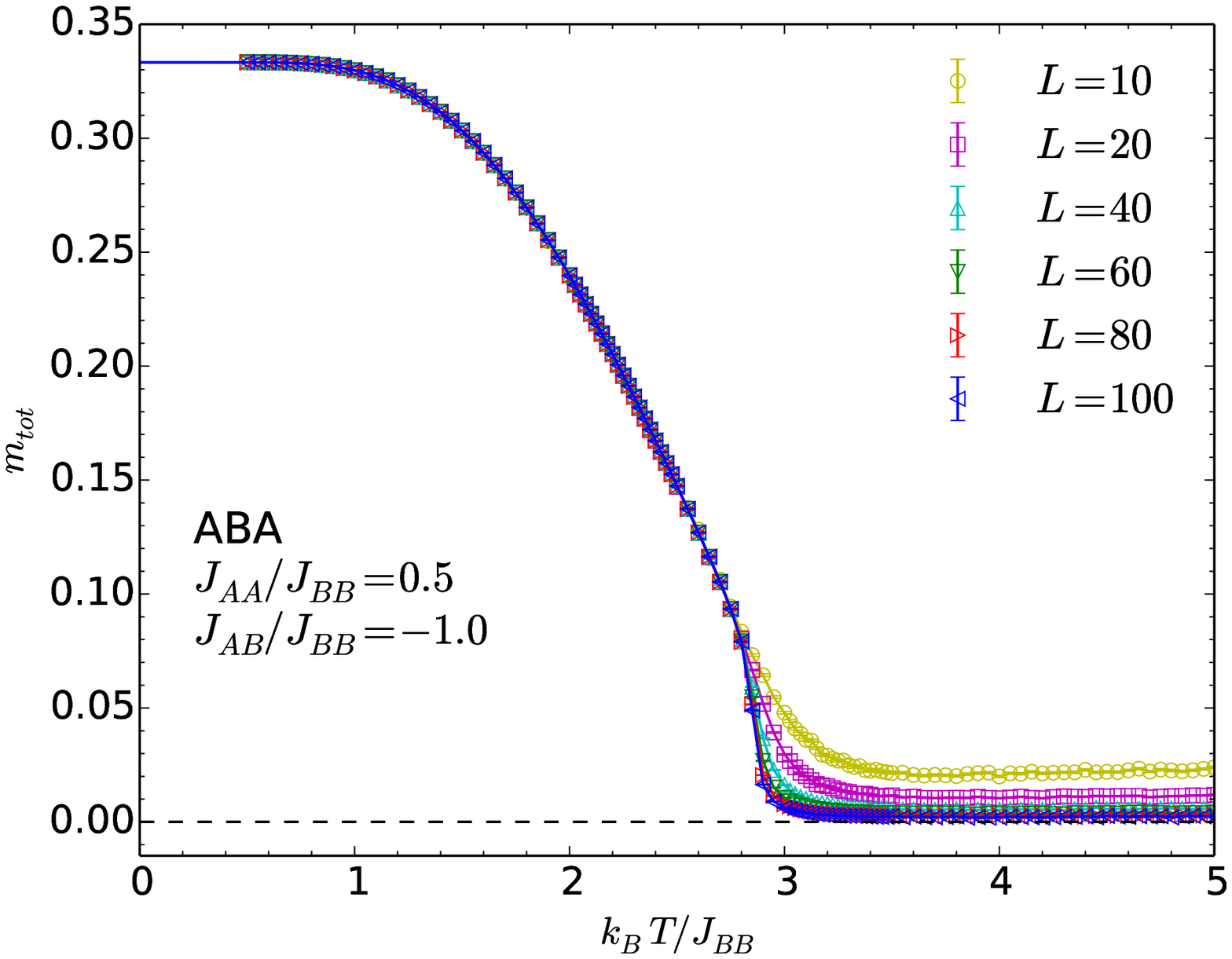}
}
\caption{
\label{fig:mags:ABA}
(Color online)
Magnetizations as functions of the dimensionless temperature $k_BT/J_{BB}$
for the \textbf{ABA} trilayer with $J_{AA}/J_{BB}=0.50$.
In (a) and (b) we show all magnetizations for $L=100$,
whereas for (c) and (d) we show only the total magnetization for several system sizes.
Figures (a) and (c), for $J_{AB}/J_{BB}=-0.1$,
show a compensation temperature $T_{comp}<T_c$ such that $m_\tot=0$.
Figures (b) and (d), for $J_{AB}/J_{BB}=-1.0$, show no compensation effect.
}
\end{center}
\end{figure}

In Figs. \ref{fig:mags:AAB} and \ref{fig:mags:ABA},
we show examples of the behavior of the magnetizations of the system
as functions of temperature for the \textbf{AAB} and \textbf{ABA} trilayers, respectively.
Both cases were obtained for $J_{AA}/J_{BB}=0.50$.
In Figs. \ref{fig:mags:AAB:a}, \ref{fig:mags:AAB:b}, \ref{fig:mags:ABA:a}, and \ref{fig:mags:ABA:b},
we show all magnetizations for $L=100$,
whereas in Figs. \ref{fig:mags:AAB:c}, \ref{fig:mags:AAB:d}, \ref{fig:mags:ABA:c}, and \ref{fig:mags:ABA:d},
we show only the total magnetization for several system sizes.
In both \textbf{AAB} and \textbf{ABA} cases,
for $J_{AB}/J_{BB}=-0.1$ we see a compensation temperature $T_{comp}<T_c$ such that $m_\tot=0$
and $m_1, m_2, m_3\neq 0$ (Figs. \ref{fig:mags:AAB:a}, \ref{fig:mags:AAB:c}, \ref{fig:mags:ABA:a}, and \ref{fig:mags:ABA:c}).
For $J_{AB}/J_{BB}=-1.0$, on the other hand, we see no compensation effect (Figs. \ref{fig:mags:AAB:b}, \ref{fig:mags:AAB:d}, \ref{fig:mags:ABA:b}, and \ref{fig:mags:ABA:d}).

For each set of values chosen for the parameters $J_{AA}/J_{BB}$, $J_{AB}/J_{BB}$, and $L$,
we performed simulations at several temperatures close to either the critical point or the compensation point.
The data generated on the simulations were analyzed
using the multiple-histogram method \cite{artigo:ferrenberg:histograma1,artigo:ferrenberg:histograma2}
in order to obtain precise estimates for $T_c$ and $T_{comp}$
as functions of the Hamiltonian parameters.
The methods used to determine the critical and compensations temperatures
are discussed in Secs. \ref{sec:tc} and \ref{sec:tcomp}, respectively.

\section{Results and Discussion}\label{sec:results}

\subsection{Determination of the critical temperatures $T_c$}\label{sec:tc}

In order to determine the critical point accurately,
we employ a finite-size scaling analysis \cite{livro:julia},
in which we examine the size dependency of certain observables measured for finite systems
of several sizes and extrapolate these results to the thermodynamic limit, i.e., $L\rightarrow\infty$.
In this approach, the singular part of the free energy density for a system of linear size $L$,
near the critical point, is assumed to obey the following scaling form:
\begin{equation}\label{eq:RGf2}
	\bar f_{\ms{sing}}(t,h,L)
	\sim L^{-d}f^0(tL^{y_t}, hL^{y_t})
\end{equation}
where $d$ is the system dimensionality, $t\equiv (T-T_c)/T_c$ is the reduced temperature,
$T_c$ is the critical temperature of the infinite system,
and $h$ is the external magnetic field given in units of $k_BT$.
The RG dimensions associated with $t$ and $h$
are $y_t=1/\nu$ and $y_h=(d+2-\eta)/2=(\gamma+\beta)/\nu$, respectively.
The critical exponents $\alpha$, $\beta$, $\gamma$ and $\nu$ are the traditional
ones associated, respectively, with the specific heat,
magnetization, magnetic susceptibility, and correlation length.

We can use Eq. \eqref{eq:RGf2} to obtain the scaling forms for other thermodynamic quantities,
e.g., for the magnetic susceptibilities at $h=0$, it reads
\begin{equation}
	\chi_\Lambda = L^{\gamma/\nu}
	\mathcal{X}_\Lambda(x_t),
	\label{FSS_chi}
\end{equation}
where $x_t\equiv tL^{1/\nu}$ is the temperature scaling variable.

As it is clear from the scaling law in Eq. \eqref{FSS_chi},
the susceptibilities diverge at the critical point only in the thermodynamic limit,
whereas for a finite system size $L$, each $\chi_\Lambda$ has a maximum at a pseudo-critical temperature $T_c(L)$,
which asymptotically approaches the real $T_c$ as $L$ increases.
This provides a powerful method to determine the critical point,
since we know that the maximum occurs when
\begin{equation}
	\left.\frac{d\mathcal{X}_\Lambda(x_t)}{dx_t}\right|_{T=T_c(L)}=0,
\end{equation}
we obtain the following relation
\begin{equation}
	\label{eq:FSS:tc}
	T_c(L) = T_c+aL^{-1/\nu},
\end{equation}
where $a$ is a constant, $T_c$ is the critical temperature and $\nu$ is the
critical exponent associated with the correlation length.

The finite-size scaling method is applicable to other quantities,
such as the specific heat and other thermodynamic derivatives \cite{artigo:landau}.
We expect the results obtained from the scaling behavior of these other quantities
to be consistent, as we were able to verify in preliminary simulations.
In this study, however, we focused only on the peak temperatures of the
magnetic susceptibilities, defined in Eq. \eqref{eq:sus},
for these peak temperatures occurred fairly close to one another and were the
sharpest peaks from all the quantities initially considered.

\begin{figure}[h]
\begin{center}
\includegraphics[width=\figwidth]{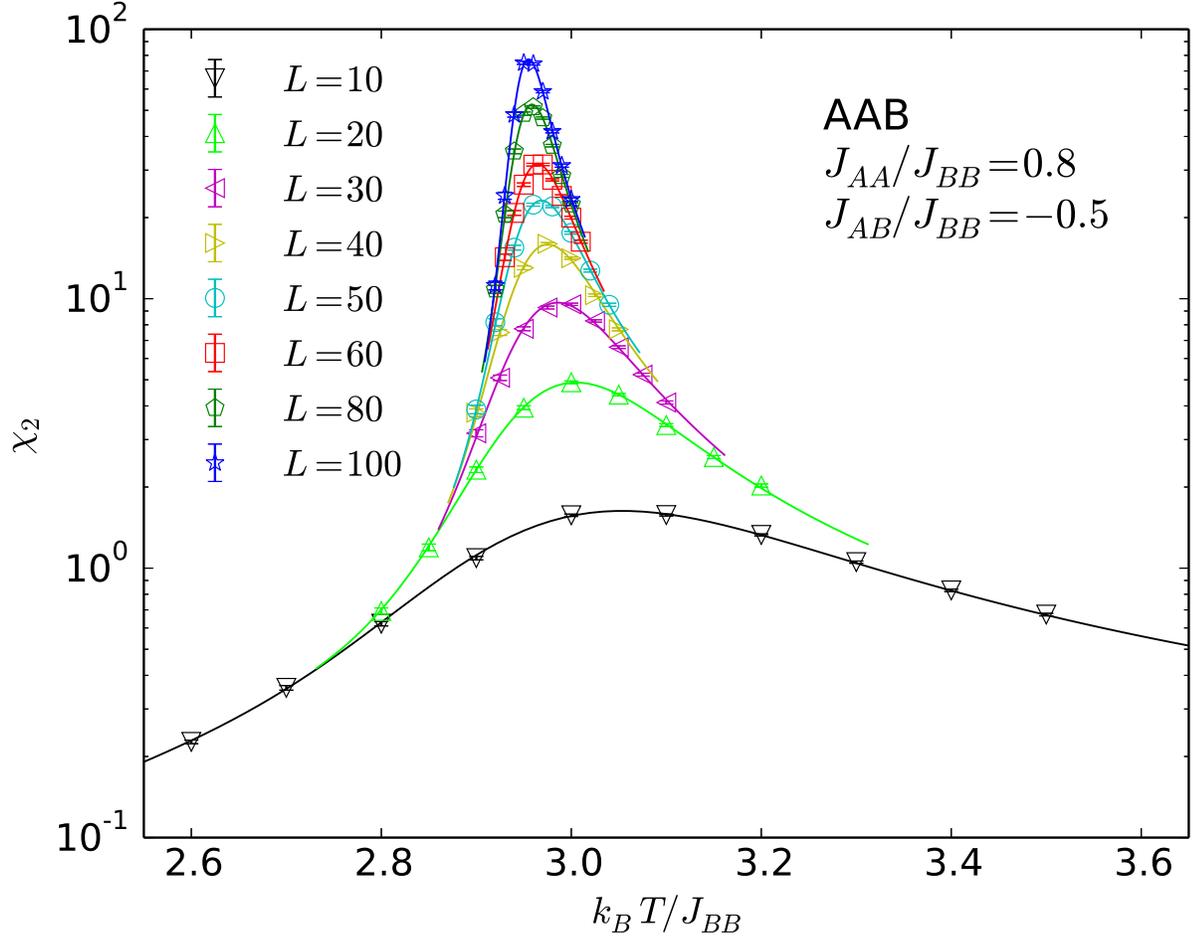}
\caption{
\label{fig:mhist:sus}
Semilog plot of the magnetic susceptibility $\chi_{2}$ as a function of the dimensionless temperature $k_BT/J_{BB}$
for an \textbf{AAB} trilayer with $J_{AA}/J_{BB}=0.80$, $J_{AB}/J_{BB}=-0.50$, and linear lattice sizes $L$ ranging from $10$ to $100$.
The symbols correspond to simulation data and the solid lines were obtained using the multiple histogram method.
Where the error bars are not visible, they are smaller than the symbols.
}
\end{center}
\end{figure}

To determine the pseudo-critical temperatures $T_c(L)$,
we carry out simulations in a temperature range that contains the peaks of the susceptibilities.
The range is typically divided in $8$ to $15$ equally spaced temperatures
and we use the multiple-histogram method to obtain $\chi_1$, $\chi_2$, $\chi_3$, and $\chi_\tot$
as continuous functions of $T$, as shown in Fig. \ref{fig:mhist:sus} for
one of the susceptibilities ($\chi_2$) of an \textbf{AAB} system
with $J_{AA}/J_{BB}=0.8$, $J_{AB}/J_{BB}=-0.5$, and $L$ from $10$ to $100$.
The location of the peak temperatures is automated using the Broyden-Fletcher-Goldfarb-Shanno (BFGS) method \cite{artigo:BFGS:1, artigo:BFGS:2}
and the errors are estimated using the blocking method \cite{livro:barkema},
i.e., we divided the data from each simulation in blocks and repeated the procedure for each block.
The errors are the standard deviation of the estimates obtained for different blocks.

The estimates of $T_c(L)$ are then used as input in Eq. \eqref{eq:FSS:tc} to perform least-square fits.
There are three free parameters in this equation to be adjusted in the fitting process,
which is feasible but requires great statistical resolution in order to produce stable and reliable estimates for all the parameters involved.
In the present work, however, we are interested only in the critical temperature and not in a precise value for the exponent $\nu$.
Thus, to avoid an unnecessary increase in computational work,
we employ the same procedure presented in Refs. \onlinecite{diaz2012feru,diaz2017monte,diaz2018monte},
in which we set a fixed value for the exponent $\nu$ and perform fits with two free parameters, instead of three.
These fits are made, for a fixed value of $\nu$, for system sizes not smaller than $L_{\smin}$
and the value of $L_\smin$ that gives the best fit is located, i.e.,
the one that minimizes the reduced weighted sum of errors, $\chi^2/n_{DOF}$,
where $n_{DOF}$ is the number of degrees of freedom.
Next, we change the values of $\nu$ and $L_\smin$ iteratively until we locate the set of values
that globally minimizes $\chi^2/n_{DOF}$ and use these values to determine our best estimate of $T_c$.

\begin{figure}[h]
\begin{center}
\includegraphics[width=\figwidth]{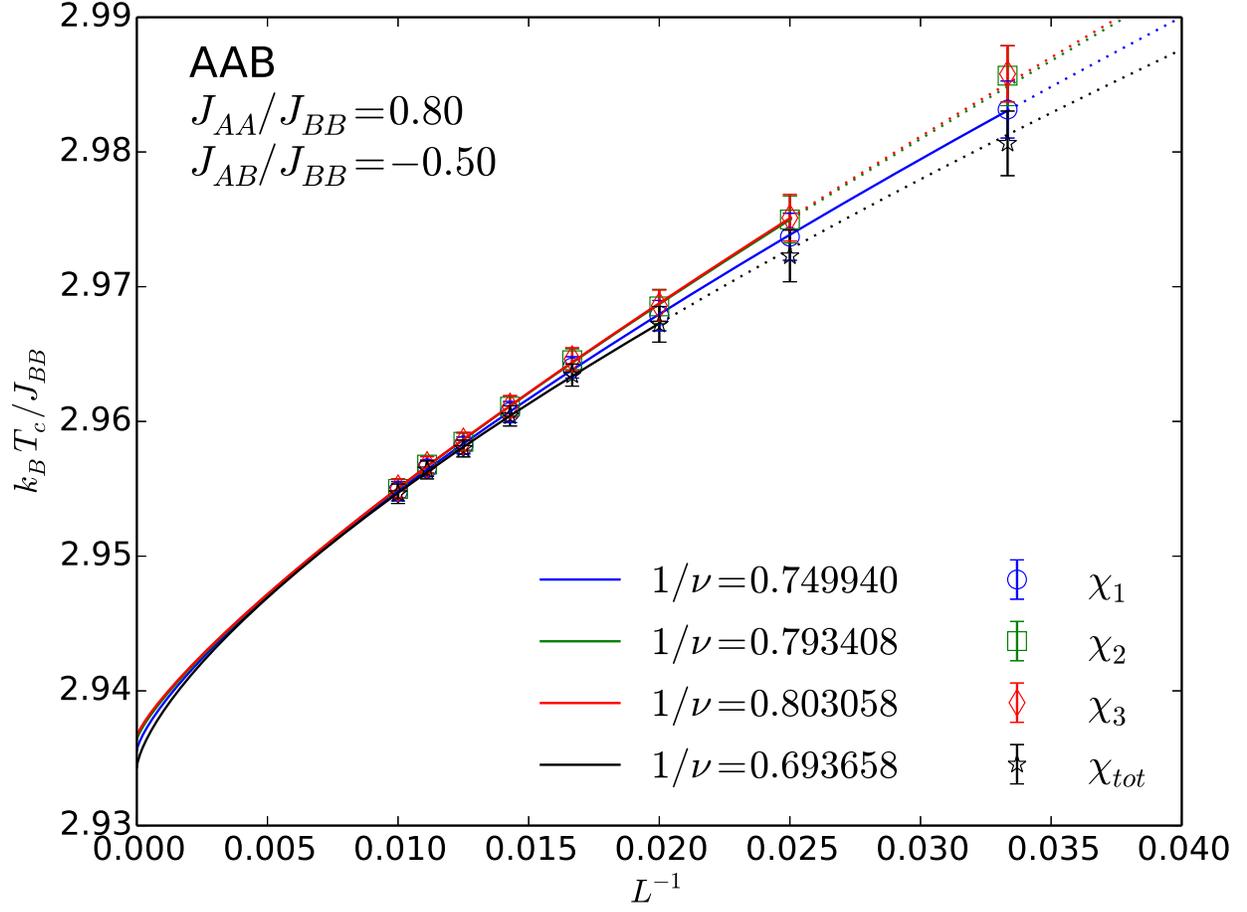}
\caption{
\label{fig:fit:tc}
Dimensionless effective critical temperature $k_BT_c(L)/J_{BB}$ as a function of $L^{-1}$
for an \textbf{AAB} trilayer with $J_{AA}/J_{BB}=0.80$ and $J_{AB}/J_{BB}=-0.50$.
The symbols correspond to $T_c(L)$ estimates made by locating the \emph{maxima} of the magnetic susceptibilities
$\chi_1$ (circles), $\chi_2$ (squares), $\chi_3$ (diamonds), and  $\chi_\tot$ (stars) for different system sizes.
The solid lines are fits performed with Eq. \eqref{eq:FSS:tc} for $L_{\smin}\leq L\leq 100$
for the values of $1/\nu$ which minimize the $\chi^2/n_{DOF}$ for each case.
The dotted lines are extrapolations of those fits for $L<L_{\smin}$.
}
\end{center}
\end{figure}

In Fig. \ref{fig:fit:tc} we show examples of fits performed with Eq. \eqref{eq:FSS:tc}
using the pseudo-critical temperatures obtained from the \emph{maxima} of the magnetic susceptibilities
for the case of an \textbf{AAB} trilayer with $J_{AA}/J_{BB}=0.80$ and $J_{AB}/J_{BB}=-0.50$.
This fits were made using the values of $L_\smin$ and $\nu$ that minimize $\chi^2/n_{DOF}$.
It is important to note that the statistical error obtained for $T_c$ through this method is small, even negligible in some cases.
Nonetheless, it is worth pointing out that this error is underestimated
when compared to the value obtained through a true non-linear fit.
Thus, to achieve a more realistic estimate for the error bar,
we follow the criterion used in Refs. \onlinecite{diaz2017monte, diaz2018monte},
in which the values obtained from fits that give $\chi^2/n_{DOF}$ up to $20\%$ larger than the minimum
are considered in the statistical analysis.

Fig. \ref{fig:fit:tc} also shows that this procedure is not adequate for a precise
determination of the values of $\nu$.
Nonetheless, as it is not our goal to obtain a precise description
of the critical behavior for the model,
we use $1/\nu$ only as an ``effective exponent'' in order to achieve a good estimate of $T_c$.
Moreover, as discussed in Refs. \onlinecite{diaz2017monte, diaz2018monte},
the final value of $T_c$ obtained through this method is not sensitive to
fluctuations around the value of $\nu$ that minimizes $\chi^2/n_{DOF}$.

\subsection{Determination of the compensation temperatures $T_{comp}$}\label{sec:tcomp}

\begin{figure}[h]
\begin{center}
\includegraphics[width=\figwidth]{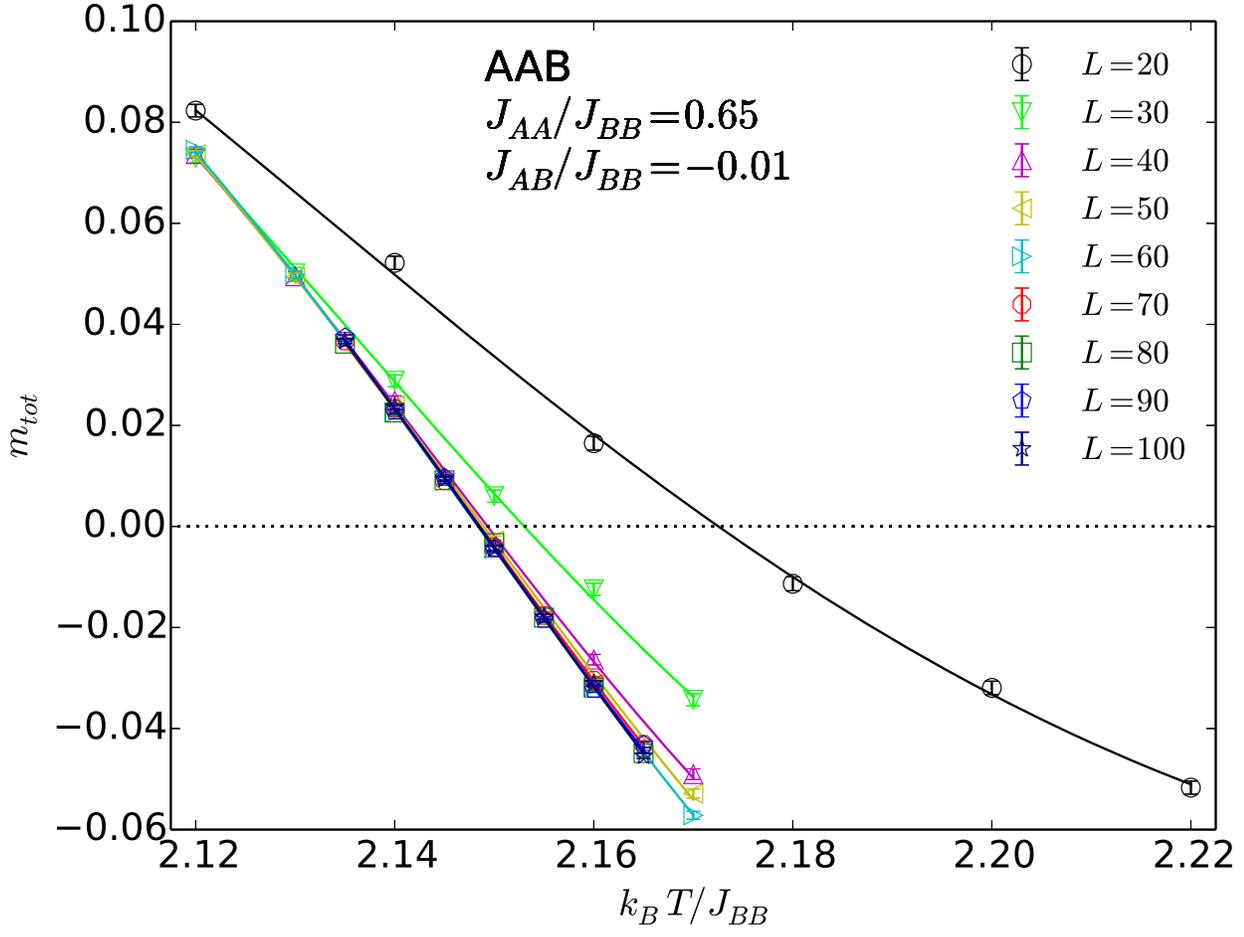}
\caption{
\label{fig:mhist:mag}
Total magnetization as a function of the dimensionless temperature $k_BT/J_{BB}$
for an \textbf{AAB} trilayer with $J_{AA}/J_{BB}=0.65$, $J_{AB}/J_{BB}=-0.01$, and linear lattice sizes $L$ ranging from $20$ to $100$.
The symbols correspond to simulation data and the solid lines were obtained using the multiple histogram method.
}
\end{center}
\end{figure}

At the compensation temperature, we have $m_\tot=0$ while
$m_1, m_2, m_3\neq 0$, as seen in Figs. \ref{fig:mags:AAB:a}, \ref{fig:mags:AAB:b}, \ref{fig:mags:ABA:a}, and \ref{fig:mags:ABA:b}.
In order to estimate this temperature we perform simulations for a range of typically $5$ to $8$ equally spaced temperatures
around $T_{comp}$ and obtain the $m_\tot$ values as a continuous function of $T$ using the multiple-histogram method,
similarly to the procedure described in Sec. \ref{sec:tc} for $T_c$.
In Fig. \ref{fig:mhist:mag} we show the total magnetization as a function of temperature
for the case of an \textbf{AAB} trilayer with $J_{AA}/J_{BB}=0.65$, $J_{AB}/J_{BB}=-0.01$,
and several system sizes $L$ from $20$ to $100$.
In this figure, the solid lines were obtained using the multiple-histogram method.
The procedure to determine the precise temperature where $m_\tot(T)=0$
for each system size is also automated and we use Brent's method \cite{artigo:brent1973}.
The error associated with $T_{comp}$ is determined via the blocking method, as discussed in Sec. \ref{sec:tc}.

\begin{figure}[h]
\begin{center}
\includegraphics[width=\figwidth]{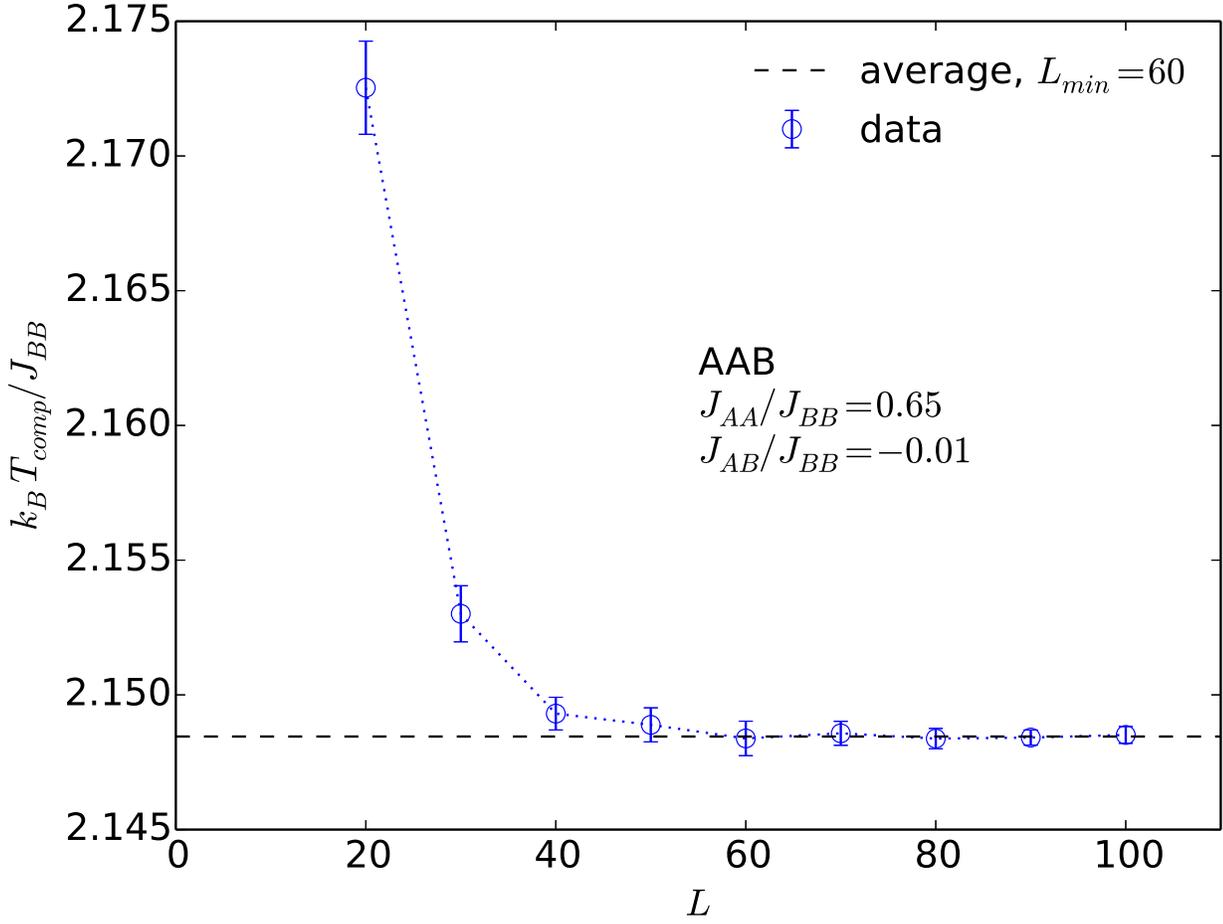}
\caption{\label{fig:fit:tcomp}
Dimensionless compensation temperature $k_BT_{comp}(L)/J_{BB}$ as
a function of linear system size $L$ for an \textbf{AAB} trilayer with $J_{AA}/J_{BB}=0.65$ and $J_{AB}/J_{BB}=-0.01$.
The symbols are estimates made by locating the zero of the total magnetization for different system sizes.
The dashed line is the average of the estimates obtained for $L\geq 60$,
which is the value that minimizes the $\chi^2/n_{DOF}$ of the fit in this particular case.
The dotted lines are connecting the symbols just to guide the eye.}
\end{center}
\end{figure}

To obtain a final estimate of $T_{comp}$, it is necessary to combine the estimates for different system sizes.
In Figs. \ref{fig:mags:AAB:c} and \ref{fig:mags:ABA:c} we see that different $T_{comp}(L)$ are fairly close to one another.
However, it is clear from Fig. \ref{fig:mhist:mag} that the smaller lattices provide somewhat inconsistent results.
Fig. \ref{fig:fit:tcomp} shows the size dependence of the compensation temperature estimates obtained from the
same data depicted in Fig. \ref{fig:mhist:mag}. We can see that, as $L$ increases,
the compensation temperature approaches a fixed value.
Thus,
we fit our data to
\begin{equation}
	\label{eq:tcomp}
	T_{comp}(L) = a = \mbox{constant},
\end{equation}
for $L\geq L_{\smin}$, which corresponds to averaging the different compensation temperatures
considering only the values of $L$ after the $T_{comp}(L)$ curve has approximately converged \cite{diaz2017monte, diaz2018monte}.
The value of $L_{\smin}$ is determined by minimizing the $\chi^2/n_{DOF}$ of the fit.
To estimate the final error bars we combine the error obtained in the fitting process
with the largest error obtained for a fixed $L$.
Note that Eq. \eqref{eq:tcomp} is consistent with the fact that the compensation phenomenon is not in any way related to criticality;
thus, we have no \emph{a priori} reason to expect a particular behavior (e.g., a power-law scaling form) for the dependence of $T_{comp}$ on $L$.

\subsection{Phase diagrams}\label{sec:phase}

Our goal in this section is to outline the contribution of each parameter
to the presence or absence of the compensation phenomenon.
To that end we determine the regions of the parameter space for which the system has
a compensation point, as seen in Figs. \ref{fig:mags:AAB:a} and \ref{fig:mags:AAB:c} for an \textbf{AAB}
trilayer and Figs. \ref{fig:mags:ABA:a} and \ref{fig:mags:ABA:c} for an \textbf{ABA} system,
and the regions for which the compensation effect does not take place,
as seen in Figs. \ref{fig:mags:AAB:b} and \ref{fig:mags:AAB:d} for an \textbf{AAB} trilayer
and Figs. \ref{fig:mags:ABA:b} and \ref{fig:mags:ABA:d} for an \textbf{ABA} system.

\begin{figure}[h]
\begin{center}
\subfigure[\textbf{AAB}.\label{fig:TvsJp:AAB}]{
\includegraphics[width=\subfigwidth]{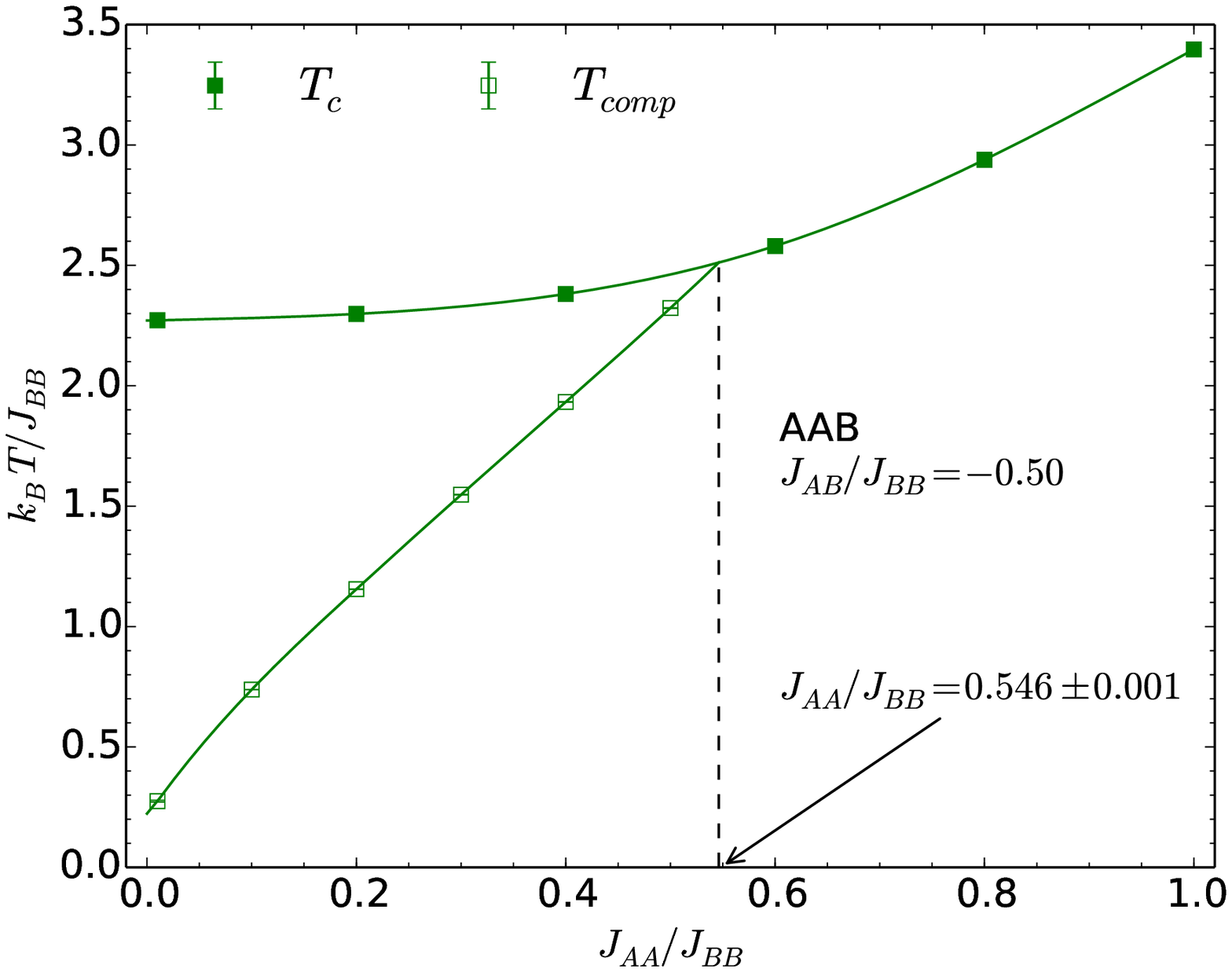}
}
\subfigure[\textbf{ABA}.\label{fig:TvsJp:ABA}]{
\includegraphics[width=\subfigwidth]{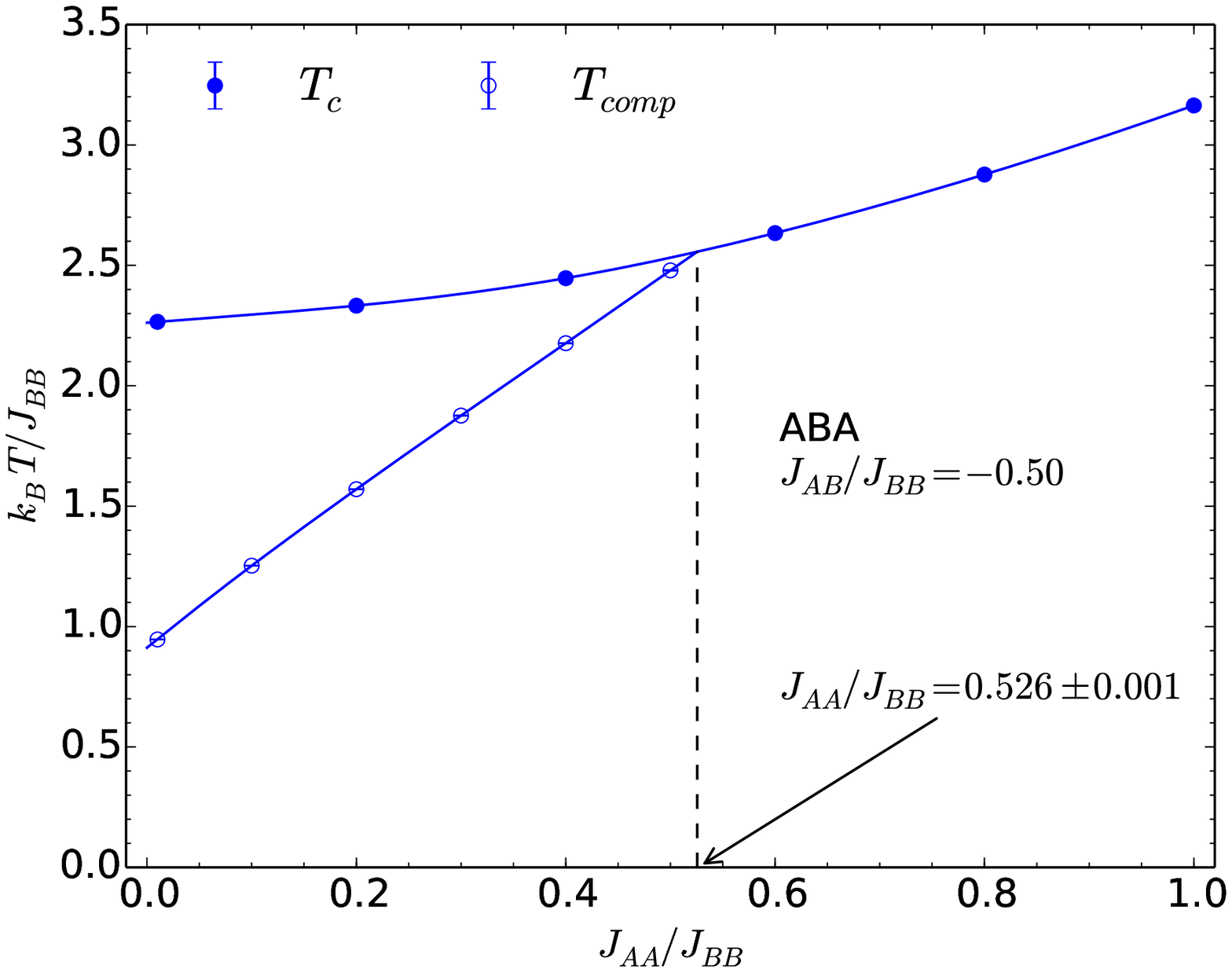}
}
\caption{\label{fig:TvsJp}
Dimensionless critical temperature $k_BT_c/J_{BB}$ (solid symbols) and compensation temperature $k_BT_{comp}/J_{BB}$ (empty symbols)
as functions of $J_{AA}/J_{BB}$ for both
(a) \textbf{AAB} and (b) \textbf{ABA} trilayers with $J_{AB}/J_{BB}=-0.50$.
The dotted lines mark the values of $J_{AA}/J_{BB}$ for which $T_{comp}=T_c$
and above which there is no compensation.}
\end{center}
\end{figure}

\begin{figure}[h]
\begin{center}
\subfigure[\textbf{AAB}.\label{fig:TvsJn:AAB}]{
\includegraphics[width=\subfigwidth]{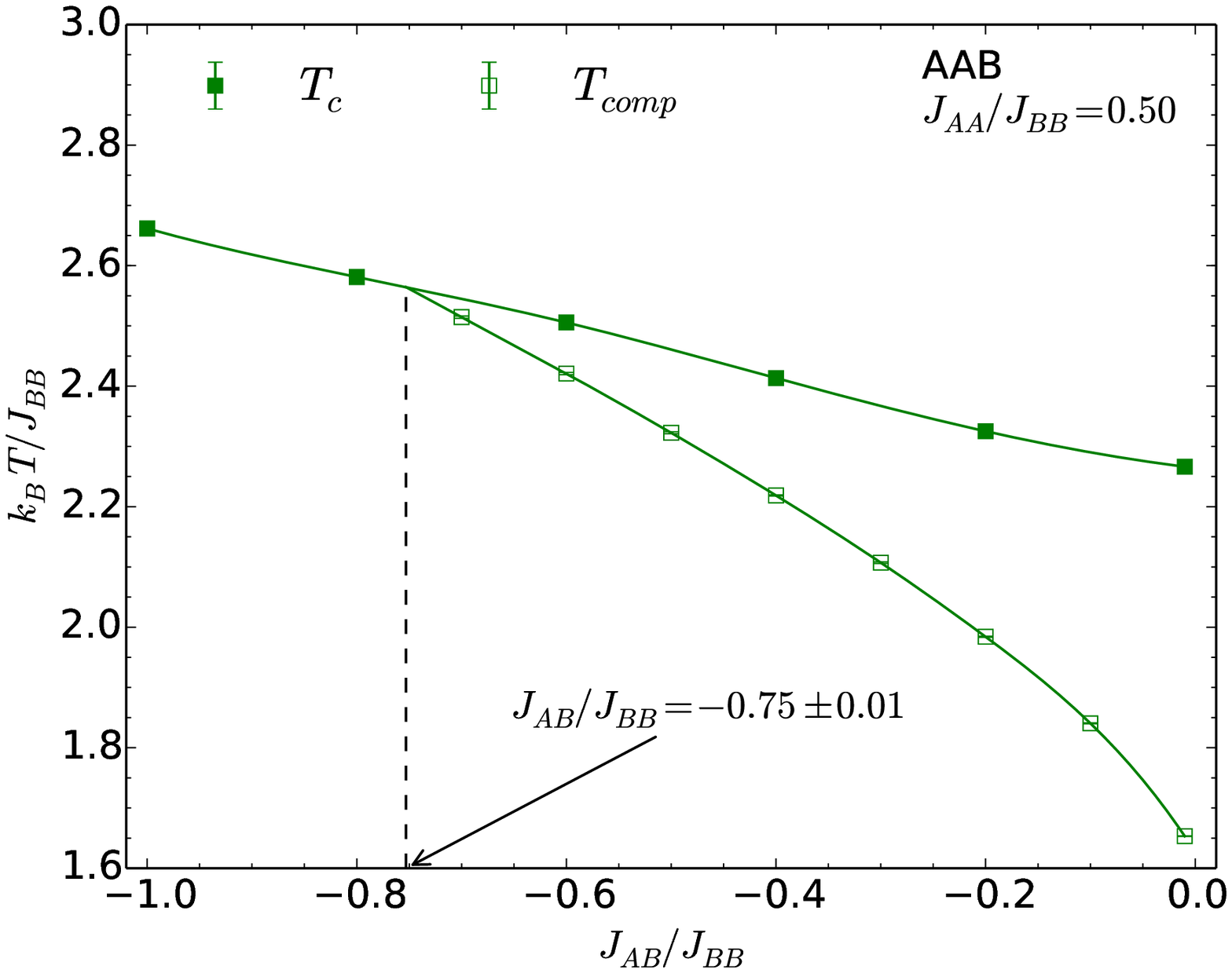}
}
\subfigure[\textbf{ABA}.\label{fig:TvsJn:ABA}]{
\includegraphics[width=\subfigwidth]{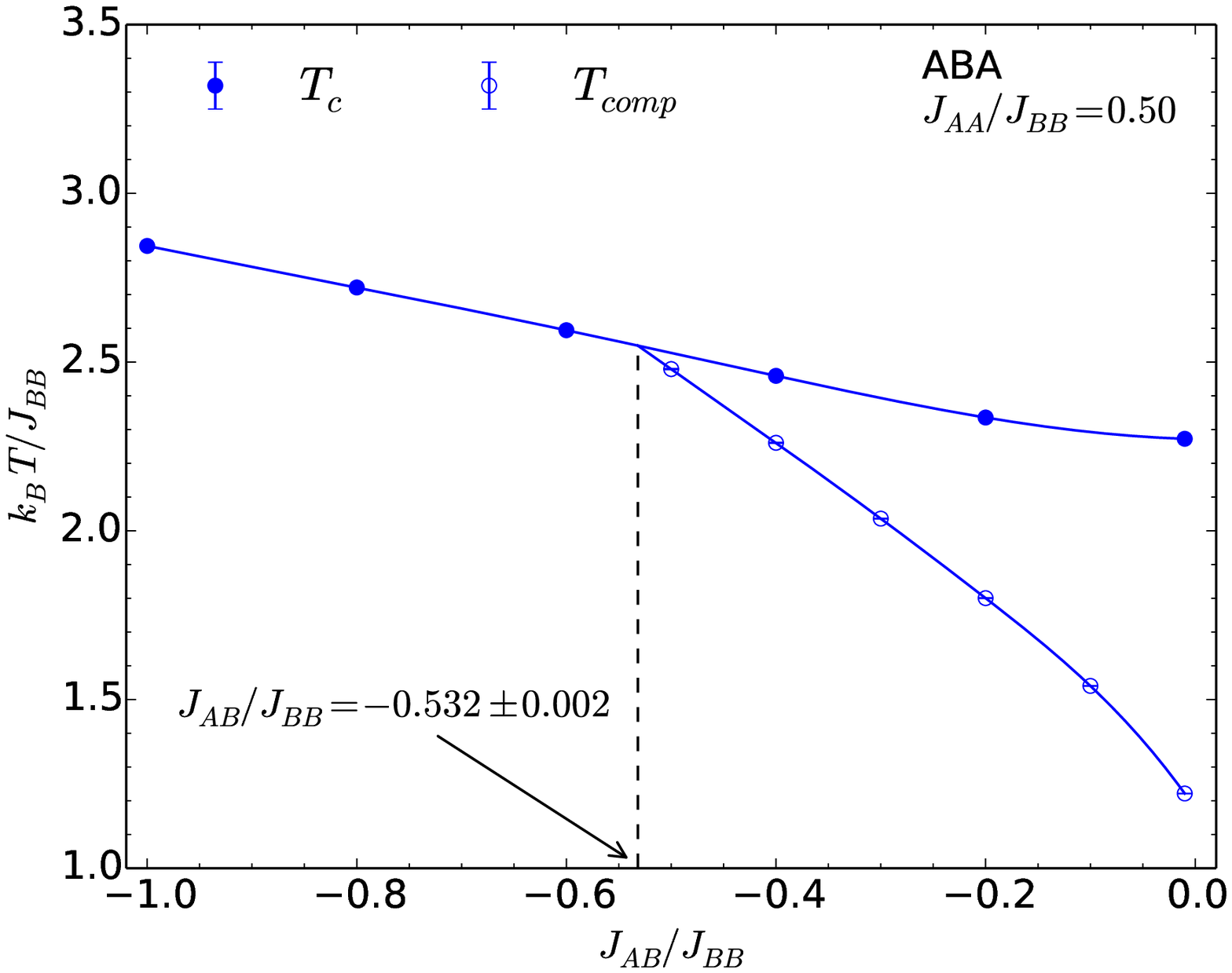}
}
\caption{\label{fig:TvsJn}
Dimensionless critical temperature $k_BT_c/J_{BB}$ (solid symbols) and compensation temperature $k_BT_{comp}/J_{BB}$ (empty symbols)
as functions of $J_{AB}/J_{BB}$ for both
(a) \textbf{AAB} and (b) \textbf{ABA} trilayers with $J_{AA}/J_{BB}=0.50$.
The dotted lines mark the values of $J_{AB}/J_{BB}$ for which $T_{comp}=T_c$
and below which there is no compensation.}
\end{center}
\end{figure}

In order to analyze the influence of $J_{AA}/J_{BB}$ in the behavior of the system,
we fix a value for $J_{AB}/J_{BB}$ and follow the procedures described in
Sec. \ref{sec:tc} and Sec. \ref{sec:tcomp} to
determine $T_c$ and $T_{comp}$ as functions of $J_{AA}/J_{BB}$,
as seen in Fig. \ref{fig:TvsJp}
for the \textbf{AAB} (Fig. \ref{fig:TvsJp:AAB}) 
and \textbf{ABA} (Fig. \ref{fig:TvsJp:ABA}) trilayers with $J_{AB}/J_{BB}=-0.50$.
In both cases, the dotted vertical lines mark the value of $J_{AA}/J_{BB}$ at which $T_c=T_{comp}$
and above which there is no compensation for each system.
Likewise, to understand the influence of $J_{AB}/J_{BB}$ in the behavior of the trilayers,
we fix a value for $J_{AA}/J_{BB}$ and obtain
$T_c$ and $T_{comp}$ as functions of $J_{AB}/J_{BB}$,
as shown in Fig. \ref{fig:TvsJn}
for both \textbf{AAB} (Fig. \ref{fig:TvsJn:AAB}) and \textbf{ABA} (Fig. \ref{fig:TvsJn:ABA})
trilayers with $J_{AB}/J_{BB}=0.50$.
The dotted vertical lines mark the value of $J_{AB}/J_{BB}$ at which $T_c=T_{comp}$
and below which there is no compensation for each system.

Our MC calculations can be compared to the EFA and MFA results reported in Ref. \onlinecite{diaz2018ferrimagnetism} for the same model.
For instance, the qualitative behavior displayed in our Fig. \ref{fig:TvsJp:AAB} agrees
with its MFA (Fig. 4(a) in Ref. \onlinecite{diaz2018ferrimagnetism}) and EFA (Fig. 4(b) in Ref. \onlinecite{diaz2018ferrimagnetism}) counterparts.
The same is true for the comparison of our Fig. \ref{fig:TvsJp:ABA}
with the analogous MFA (Fig. 5(a) in Ref. \onlinecite{diaz2018ferrimagnetism}) and EFA (Fig. 5(b) in Ref. \onlinecite{diaz2018ferrimagnetism}) ones.
Nonetheless, the quantitative results are significantly different in all cases.
Namely, in the same way the EFA values for the critical temperatures are consistently lower than those for the MFA \cite{diaz2018ferrimagnetism},
the same is true for the MC estimates, which are lower than both the EFA and MFA ones.

Comparing the $T_c$ values in our Fig. \ref{fig:TvsJp:AAB} and those from Fig. 4 in Ref. \onlinecite{diaz2018ferrimagnetism},
we can see that, for an \textbf{AAB} trilayer with $J_{AB}/J_{BB}=-0.5$, the EFA estimates range from $24.0\%$ to $37.5\%$ higher than the MC ones,
whereas the MFA estimates are from $50.7\%$ to $78.8\%$ higher than the MC estimate.
In both cases the largest discrepancy occurs for small $J_{AA}/J_{BB}$ values.
On the other hand, for $T_{comp}$ this discrepancies are much smaller,
being less than $1\%$ in both the EFA and MFA for small $J_{AA}/J_{BB}$ and increasing as $T_{comp}$ approaches $T_c$ at higher $J_{AA}/J_{BB}$ values.
For $J_{AA}/J_{BB}=0.5$ for example,
the percentile deviation between the MFA and MC compensation temperatures is $18.0\%$
while the EFA estimate is only $3.8\%$ larger than its MC counterpart.

This is expected since both the mean-field and the effective-field approximations
neglect spin-spin correlations that are fully taken into account in Monte Carlo simulations.
Therefore, both MFA and EFA approaches overestimate the critical temperatures,
whereas MC simulations provide $T_c$ estimates that are
much closer to the true values than their mean-field-like counterparts.
Since the effective-field approach still takes into account short-range correlations,
which are entirely neglected by a standard MFA,
the EFA temperatures should still be closer to the MC values than the MFA ones.
It is worth stressing that the same occurs when we contrast
pair approximation (PA) \cite{balcerzak2014ferrimagnetism} and Monte Carlo \cite{diaz2017monte} results
for a site-diluted Ising bilayer,
in which case the PA temperatures are higher than the MC ones.
Although the PA takes into account longer-range correlations than both EFA and MFA,
it still systematically overestimates the temperatures since it is a mean-field-like approximation.
Thus, we would expect a PA study of the trilayer systems presented in this work
to provide $T_c$ estimates that are between the MC and EFA values for each set of Hamiltonian parameters.

Fig. \ref{fig:TvsJn} further helps to highlight the differences between MFA, EFA, and MC results.
For $J_{AA}/J_{BB}=0.50$,
our MC simulations show that there is no compensation below $J_{AB}/J_{BB}=-0.75\pm 0.01$
and $J_{AB}/J_{BB}=-0.532\pm 0.002$ for the \textbf{AAB} and \textbf{ABA} trilayers, respectively.
On the other hand, the mean-field and effective-field approximations
predict that both types of trilayer will be in a ferrimagnetic phase with compensation
for $J_{AA}/J_{BB}=0.50$, irrespective of the value of $J_{AB}/J_{BB}$,
as it is clear from Fig. 8 in Ref. \onlinecite{diaz2018ferrimagnetism}.

Regarding the compensation temperature estimates obtained through different approximations,
though, it is clear from the comparison between our Fig. \ref{fig:TvsJp}
and Figs. 4 and 5 in Ref. \onlinecite{diaz2018ferrimagnetism},
that there are no drastic differences between the MC, EFA, and MFA values.
However, since the $T_c$ estimates are systematically different for the approximations considered,
as discussed in the last paragraph, we expect a significant change in the values of $J_{AA}/J_{BB}$
for which the $T_c$ and $T_{comp}$ curves intersect,
which we shall henceforth call $(J_{AA}/J_{BB})^\ast$ for convenience.
And in fact, for the \textbf{AAB} system, we have
$(J_{AA}/J_{BB})^\ast=0.762381$, $(J_{AA}/J_{BB})^\ast=0.702061$, and $(J_{AA}/J_{BB})^\ast=0.546\pm 0.001$
for the MFA, EFA, and MC approaches, respectively.
Similarly, for the \textbf{ABA} trilayer, we have
$(J_{AA}/J_{BB})^\ast=0.875053$, $(J_{AA}/J_{BB})^\ast=0.796088$, and $(J_{AA}/J_{BB})^\ast=0.526\pm 0.001$
for the MFA, EFA, and MC approaches, respectively.
This is a clear indication that the area of the region occupied by a ferrimagnetic phase with compensation
is overestimated by the mean-field-like approximations, and that this area
decreases as we increase the complexity of the approximation used.

\begin{figure}[h]
\begin{center}
\includegraphics[width=\figwidth]{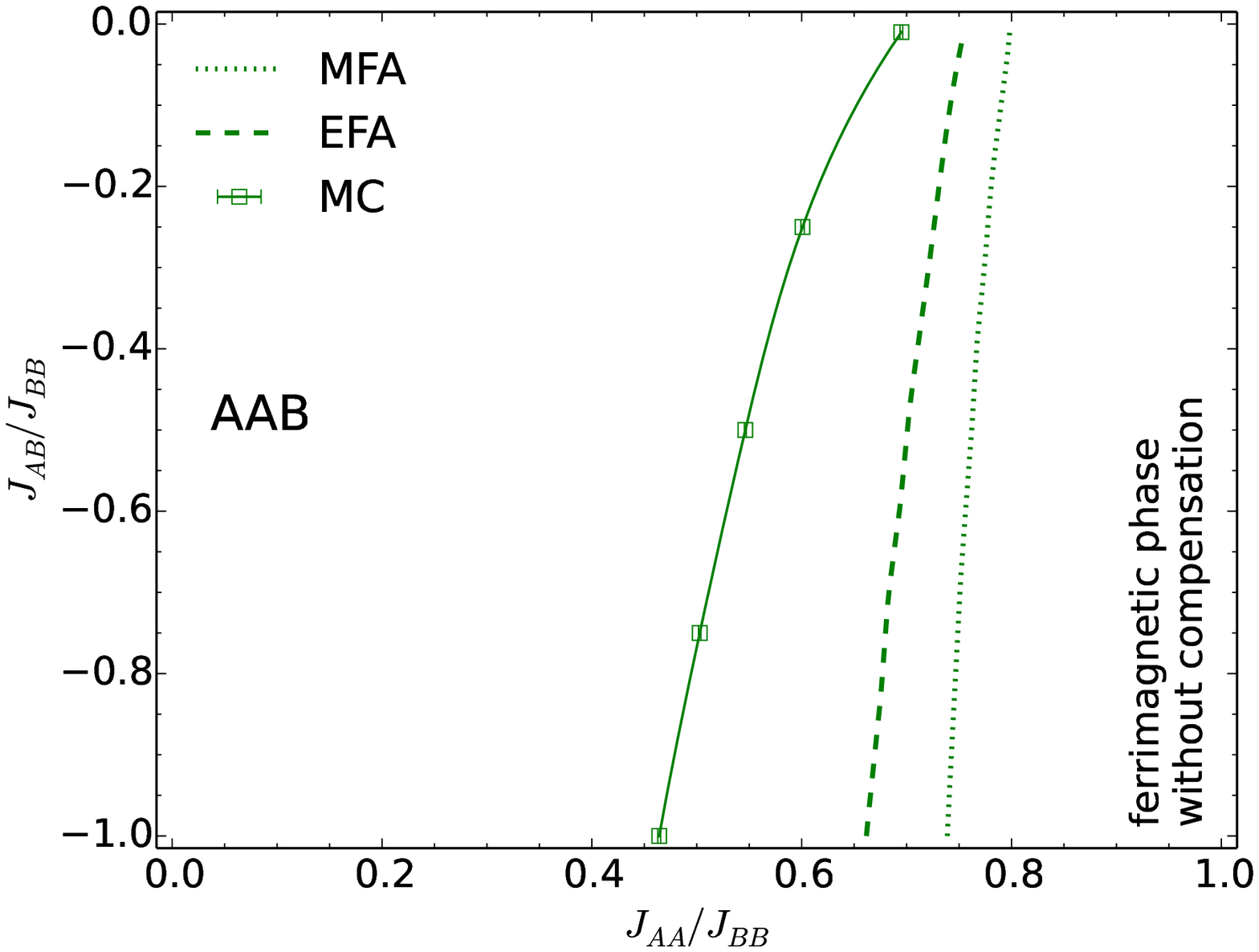}
\caption{\label{fig:phase:AAB}
Phase diagrams for the \textbf{AAB} trilayer.
The squares were obtained through MC simulations and the solid lines are either cubic spline interpolations
or linear extrapolations just to guide the eye.
The effective-field approximation (dashed line) and mean-field approximation (dotted line)
results correspond to those shown in Fig. 8 of Ref. \onlinecite{diaz2018ferrimagnetism}
and are reproduced here for comparison purposes only.
In all cases, the lines mark the separation between a ferrimagnetic phase with compensation (to the left)
and a ferrimagnetic phase without compensation (to the right).}
\end{center}
\end{figure}

\begin{figure}[h]
\begin{center}
\includegraphics[width=\figwidth]{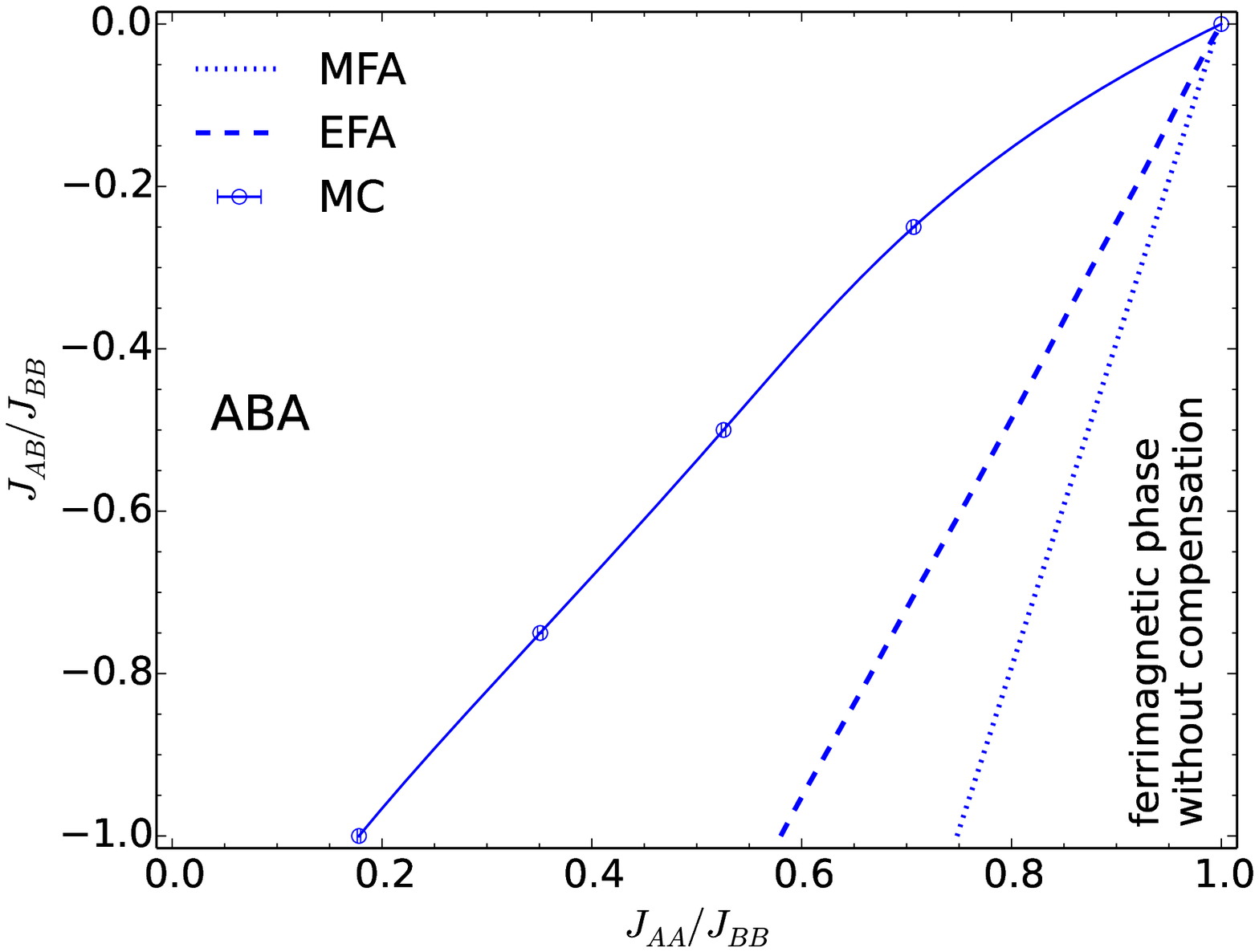}
\caption{\label{fig:phase:ABA}
Phase diagrams for the \textbf{ABA} trilayer.
The circles were obtained through MC simulations and the solid lines are either cubic spline interpolations
or linear extrapolations just to guide the eye.
The effective-field approximation (dashed line) and mean-field approximation (dotted line)
results correspond to those shown in Fig. 8 of Ref. \onlinecite{diaz2018ferrimagnetism}
and are reproduced here for comparison purposes only.
In all cases, the lines mark the separation between a ferrimagnetic phase with compensation (to the left)
and a ferrimagnetic phase without compensation (to the right).}
\end{center}
\end{figure}

If we repeat the procedure used to obtain Fig. \ref{fig:TvsJp} for other values of $J_{AB}/J_{BB}$,
we can obtain a phase diagram dividing the parameter space of our Hamiltonian in two distinct regions of interest.
One is a ferrimagnetic phase for which there is no compensation at any temperature
and the second is a ferrimagnetic phase where there is a compensation point at a certain temperature $T_{comp}$.
We present these results in Fig. \ref{fig:phase:AAB} for the \textbf{AAB} trilayer.
In this figure we also reproduce the MFA and EFA results reported in Ref. \onlinecite{diaz2018ferrimagnetism}
for comparison purposes only.
In all cases, the lines mark the separation between a ferrimagnetic phase with compensation (to the left)
and a ferrimagnetic phase without compensation (to the right).
Analogously, in Fig. \ref{fig:phase:ABA} we present the MC, EFA, and MFA phase diagrams for the \textbf{ABA} system.
These diagrams show that, in both trilayer types,
there is always a compensation temperature for a sufficiently small $J_{AA}/J_{BB}$
irrespective of the value of $J_{AB}/J_{BB}$,
although the range of values of $J_{AA}/J_{BB}$ for which the phenomenon occurs increases
as the \textbf{A-B} interplanar coupling gets weaker.
This behavior is consistent for MC, EFA, and MFA approaches
and similar to what is reported for the diluted bilayer \cite{balcerzak2014ferrimagnetism, diaz2017monte}
and multilayer \cite{szalowski2014normal, diaz2018monte} systems for sufficiently small dilutions.

For the three approximations considered,
the main difference we see when contrasting the behaviors depicted in Figs. \ref{fig:phase:AAB} and \ref{fig:phase:ABA}
is that the lines separating the phases are closer to straight vertical lines
for the \textbf{AAB} trilayer than for the \textbf{ABA} system,
i.e., the value of $(J_{AA}/J_{BB})^\ast$ is less sensitive to
the value of $J_{AB}/J_{BB}$ for the former system than for the latter.
This is consistent with the fact that the number of \textbf{A}-\textbf{B} bonds
in the \textbf{AAB} trilayer is only half that of the \textbf{ABA} system.
In addition, Figs. \ref{fig:phase:AAB} and \ref{fig:phase:ABA}
show that the area occupied by the ferrimagnetic phase with compensation in the $J_{AB}\times J_{AA}$ diagrams
is the smallest for the MC approach, followed by the EFA, and finally by the MFA.
This happens for both types of trilayer and confirms the trend seen when comparing
Fig. \ref{fig:TvsJp} with the results reported in Ref. \onlinecite{diaz2018ferrimagnetism}.
We see the same behavior when we contrast the
PA \cite{balcerzak2014ferrimagnetism} and MC \cite{diaz2017monte} results
for the Ising bilayer, in which case the smaller area is also obtained through Monte Carlo simulations,
i.e., the area seems to decrease as we use more accurate approximations.
Thus, we expect that if the pair approximation were applied to the trilayer systems,
the line separating the phases with and without compensation would fall
in between the dashed (EFA) and solid (MC) lines in both Figs. \ref{fig:phase:AAB} (\textbf{AAB}) and \ref{fig:phase:ABA} (\textbf{ABA}).

\section{Conclusion}\label{sec:conclusion}

In summary, we have investigated the magnetic and thermodynamic properties of a spin-$1/2$ Ising trilayer.
The system is composed of three planes, each of which can only have atoms of one out of two types (\textbf{A} or \textbf{B}).
The interactions between pairs of atoms of the same type (\textbf{A}-\textbf{A} or \textbf{B}-\textbf{B} bonds) are ferromagnetic
while the interactions between pairs of atoms of different types (\textbf{A}-\textbf{B} bonds) are antiferromagnetic.
The study is carried out in a Monte Carlo approach, aided by
a multiple histogram reweighting technique and finite-size scaling methods.
We verified the occurrence of a compensation phenomenon
and determined the compensation temperatures, as well as the critical temperatures of the model,
for a range of values of the Hamiltonian parameters.

We present phase diagrams and a detailed discussion
about the conditions for the occurrence of the compensation phenomenon.
For instance, we see that the phenomenon is only possible if $J_{AA}<J_{BB}$
and that the range of values of $J_{AA}/J_{BB}$ for which there is compensation
increases as $|J_{AB}/J_{BB}|$ gets smaller,
as it is also the case for similar systems containing a mixture of ferromagnetic and
antiferromagnetic bonds \cite{balcerzak2014ferrimagnetism, szalowski2014normal, diaz2017monte}.
The summary of the results is presented in a convenient way on $J_{AB}\times J_{AA}$
diagrams which separate the Hamiltonian parameter-space
in two distinct regions: one corresponding to a ferrimagnetic phase where
the system has a compensation point and the other
is a ferrimagnetic phase without compensation.

We compare our results with both mean-field and effective-field approximations
applied to the same model \cite{diaz2018ferrimagnetism}
and we confirm that the compensation phenomenon is robust and
occurs for all values of \textbf{A-B} exchanges in the range $-1.0\leq J_{AB}/J_{BB}<0.0$.
Although it is clear from this comparison that the area of the parameter space occupied by the
ferrimagnetic phase with compensation diminishes as we increase the accuracy of our approximation,
the area obtained in this work through MC simulations is still fairly large
if compared with the results for the diluted Ising bilayer \cite{diaz2017monte}
and multilayer \cite{diaz2018monte} systems, especially as dilution is increased.
It is worth stressing that our results show relevant quantitative differences with those
obtained from MFA and EFA.
Therefore, the MC results may be an important tool for experimentalists interested in
building layered materials with a priori desired physical properties.



\begin{acknowledgments}
We are indebted to Prof. Dr. Lucas Nicolao for suggestions and helpful discussions.
We would also like to thank Sigrid Anja Reichert for a critical reading of the manuscript.
This work has been partially supported by the Brazilian Agency CNPq.
\end{acknowledgments}


 \newcommand{\noop}[1]{}
%

\newpage

\begin{thebibliography}{66}%
\makeatletter
\providecommand \@ifxundefined [1]{%
 \@ifx{#1\undefined}
}%
\providecommand \@ifnum [1]{%
 \ifnum #1\expandafter \@firstoftwo
 \else \expandafter \@secondoftwo
 \fi
}%
\providecommand \@ifx [1]{%
 \ifx #1\expandafter \@firstoftwo
 \else \expandafter \@secondoftwo
 \fi
}%
\providecommand \natexlab [1]{#1}%
\providecommand \enquote  [1]{``#1''}%
\providecommand \bibnamefont  [1]{#1}%
\providecommand \bibfnamefont [1]{#1}%
\providecommand \citenamefont [1]{#1}%
\providecommand \href@noop [0]{\@secondoftwo}%
\providecommand \href [0]{\begingroup \@sanitize@url \@href}%
\providecommand \@href[1]{\@@startlink{#1}\@@href}%
\providecommand \@@href[1]{\endgroup#1\@@endlink}%
\providecommand \@sanitize@url [0]{\catcode `\\12\catcode `\$12\catcode
  `\&12\catcode `\#12\catcode `\^12\catcode `\_12\catcode `\%12\relax}%
\providecommand \@@startlink[1]{}%
\providecommand \@@endlink[0]{}%
\providecommand \url  [0]{\begingroup\@sanitize@url \@url }%
\providecommand \@url [1]{\endgroup\@href {#1}{\urlprefix }}%
\providecommand \urlprefix  [0]{URL }%
\providecommand \Eprint [0]{\href }%
\providecommand \doibase [0]{http://dx.doi.org/}%
\providecommand \selectlanguage [0]{\@gobble}%
\providecommand \bibinfo  [0]{\@secondoftwo}%
\providecommand \bibfield  [0]{\@secondoftwo}%
\providecommand \translation [1]{[#1]}%
\providecommand \BibitemOpen [0]{}%
\providecommand \bibitemStop [0]{}%
\providecommand \bibitemNoStop [0]{.\EOS\space}%
\providecommand \EOS [0]{\spacefactor3000\relax}%
\providecommand \BibitemShut  [1]{\csname bibitem#1\endcsname}%
\let\auto@bib@innerbib\@empty
\bibitem [{\citenamefont {Connell}\ \emph {et~al.}(1982)\citenamefont
  {Connell}, \citenamefont {Allen},\ and\ \citenamefont
  {Mansuripur}}]{connell1982magneto}%
  \BibitemOpen
  \bibfield  {author} {\bibinfo {author} {\bibfnamefont {G.}~\bibnamefont
  {Connell}}, \bibinfo {author} {\bibfnamefont {R.}~\bibnamefont {Allen}}, \
  and\ \bibinfo {author} {\bibfnamefont {M.}~\bibnamefont {Mansuripur}},\
  }\href@noop {} {\bibfield  {journal} {\bibinfo  {journal} {Journal of Applied
  Physics}\ }\textbf {\bibinfo {volume} {53}},\ \bibinfo {pages} {7759}
  (\bibinfo {year} {1982})}\BibitemShut {NoStop}%
\bibitem [{\citenamefont {Gr\"unberg}\ \emph {et~al.}(1986)\citenamefont
  {Gr\"unberg}, \citenamefont {Schreiber}, \citenamefont {Pang}, \citenamefont
  {Brodsky},\ and\ \citenamefont {Sowers}}]{grumberg1986layered}%
  \BibitemOpen
  \bibfield  {author} {\bibinfo {author} {\bibfnamefont {P.}~\bibnamefont
  {Gr\"unberg}}, \bibinfo {author} {\bibfnamefont {R.}~\bibnamefont
  {Schreiber}}, \bibinfo {author} {\bibfnamefont {Y.}~\bibnamefont {Pang}},
  \bibinfo {author} {\bibfnamefont {M.~B.}\ \bibnamefont {Brodsky}}, \ and\
  \bibinfo {author} {\bibfnamefont {H.}~\bibnamefont {Sowers}},\ }\href
  {\doibase 10.1103/PhysRevLett.57.2442} {\bibfield  {journal} {\bibinfo
  {journal} {Phys. Rev. Lett.}\ }\textbf {\bibinfo {volume} {57}},\ \bibinfo
  {pages} {2442} (\bibinfo {year} {1986})}\BibitemShut {NoStop}%
\bibitem [{\citenamefont {Camley}\ and\ \citenamefont
  {Barna{\'s}}(1989)}]{camley1989theory}%
  \BibitemOpen
  \bibfield  {author} {\bibinfo {author} {\bibfnamefont {R.~E.}\ \bibnamefont
  {Camley}}\ and\ \bibinfo {author} {\bibfnamefont {J.}~\bibnamefont
  {Barna{\'s}}},\ }\href@noop {} {\bibfield  {journal} {\bibinfo  {journal}
  {Physical Review Letters}\ }\textbf {\bibinfo {volume} {63}},\ \bibinfo
  {pages} {664} (\bibinfo {year} {1989})}\BibitemShut {NoStop}%
\bibitem [{\citenamefont {Phan}\ and\ \citenamefont
  {Yu}(2007)}]{phan2007review}%
  \BibitemOpen
  \bibfield  {author} {\bibinfo {author} {\bibfnamefont {M.-H.}\ \bibnamefont
  {Phan}}\ and\ \bibinfo {author} {\bibfnamefont {S.-C.}\ \bibnamefont {Yu}},\
  }\href@noop {} {\bibfield  {journal} {\bibinfo  {journal} {Journal of
  Magnetism and Magnetic Materials}\ }\textbf {\bibinfo {volume} {308}},\
  \bibinfo {pages} {325} (\bibinfo {year} {2007})}\BibitemShut {NoStop}%
\bibitem [{\citenamefont {Cullity}\ and\ \citenamefont
  {Graham}(2008)}]{cullity2011introduction}%
  \BibitemOpen
  \bibfield  {author} {\bibinfo {author} {\bibfnamefont {B.~D.}\ \bibnamefont
  {Cullity}}\ and\ \bibinfo {author} {\bibfnamefont {C.~D.}\ \bibnamefont
  {Graham}},\ }\href@noop {} {\emph {\bibinfo {title} {Introduction to magnetic
  materials}}},\ \bibinfo {edition} {2nd}\ ed.\ (\bibinfo  {publisher} {John
  Wiley \& Sons},\ \bibinfo {address} {New Jersey, USA},\ \bibinfo {year}
  {2008})\BibitemShut {NoStop}%
\bibitem [{\citenamefont {Shieh}\ and\ \citenamefont
  {Kryder}(1986)}]{shieh1986magneto}%
  \BibitemOpen
  \bibfield  {author} {\bibinfo {author} {\bibfnamefont {H.-P.~D.}\
  \bibnamefont {Shieh}}\ and\ \bibinfo {author} {\bibfnamefont {M.~H.}\
  \bibnamefont {Kryder}},\ }\href@noop {} {\bibfield  {journal} {\bibinfo
  {journal} {Applied physics letters}\ }\textbf {\bibinfo {volume} {49}},\
  \bibinfo {pages} {473} (\bibinfo {year} {1986})}\BibitemShut {NoStop}%
\bibitem [{\citenamefont {Ostorero}\ \emph {et~al.}(1994)\citenamefont
  {Ostorero}, \citenamefont {Escorne}, \citenamefont {Pecheron-Guegan},
  \citenamefont {Soulette},\ and\ \citenamefont {Le~Gall}}]{ostorero1994dy}%
  \BibitemOpen
  \bibfield  {author} {\bibinfo {author} {\bibfnamefont {J.}~\bibnamefont
  {Ostorero}}, \bibinfo {author} {\bibfnamefont {M.}~\bibnamefont {Escorne}},
  \bibinfo {author} {\bibfnamefont {A.}~\bibnamefont {Pecheron-Guegan}},
  \bibinfo {author} {\bibfnamefont {F.}~\bibnamefont {Soulette}}, \ and\
  \bibinfo {author} {\bibfnamefont {H.}~\bibnamefont {Le~Gall}},\ }\href@noop
  {} {\bibfield  {journal} {\bibinfo  {journal} {Journal of Applied Physics}\
  }\textbf {\bibinfo {volume} {75}},\ \bibinfo {pages} {6103} (\bibinfo {year}
  {1994})}\BibitemShut {NoStop}%
\bibitem [{\citenamefont {Herman}\ and\ \citenamefont
  {Sitter}(2012)}]{herman2012molecular}%
  \BibitemOpen
  \bibfield  {author} {\bibinfo {author} {\bibfnamefont {M.~A.}\ \bibnamefont
  {Herman}}\ and\ \bibinfo {author} {\bibfnamefont {H.}~\bibnamefont
  {Sitter}},\ }\href@noop {} {\emph {\bibinfo {title} {Molecular beam epitaxy:
  fundamentals and current status}}},\ Vol.~\bibinfo {volume} {7}\ (\bibinfo
  {publisher} {Springer Science \& Business Media},\ \bibinfo {year}
  {2012})\BibitemShut {NoStop}%
\bibitem [{\citenamefont
  {Stringfellow}(1999)}]{stringfellow1999organometallic}%
  \BibitemOpen
  \bibfield  {author} {\bibinfo {author} {\bibfnamefont {G.~B.}\ \bibnamefont
  {Stringfellow}},\ }\href@noop {} {\emph {\bibinfo {title} {Organometallic
  vapor-phase epitaxy: theory and practice}}}\ (\bibinfo  {publisher} {Academic
  Press},\ \bibinfo {year} {1999})\BibitemShut {NoStop}%
\bibitem [{\citenamefont {Singh}\ and\ \citenamefont
  {Narayan}(1990)}]{singh1990pulsed}%
  \BibitemOpen
  \bibfield  {author} {\bibinfo {author} {\bibfnamefont {R.~K.}\ \bibnamefont
  {Singh}}\ and\ \bibinfo {author} {\bibfnamefont {J.}~\bibnamefont
  {Narayan}},\ }\href {\doibase 10.1103/PhysRevB.41.8843} {\bibfield  {journal}
  {\bibinfo  {journal} {Phys. Rev. B}\ }\textbf {\bibinfo {volume} {41}},\
  \bibinfo {pages} {8843} (\bibinfo {year} {1990})}\BibitemShut {NoStop}%
\bibitem [{\citenamefont {Chrisey}\ and\ \citenamefont
  {Hubler}(1994)}]{chrisey1994pulsed}%
  \BibitemOpen
  \bibfield  {author} {\bibinfo {author} {\bibfnamefont {D.~B.}\ \bibnamefont
  {Chrisey}}\ and\ \bibinfo {author} {\bibfnamefont {G.~K.}\ \bibnamefont
  {Hubler}},\ }\href@noop {} {\  (\bibinfo {year} {1994})}\BibitemShut
  {NoStop}%
\bibitem [{\citenamefont {Leskel{\"a}}\ and\ \citenamefont
  {Ritala}(2003)}]{leskela2003atomic}%
  \BibitemOpen
  \bibfield  {author} {\bibinfo {author} {\bibfnamefont {M.}~\bibnamefont
  {Leskel{\"a}}}\ and\ \bibinfo {author} {\bibfnamefont {M.}~\bibnamefont
  {Ritala}},\ }\href@noop {} {\bibfield  {journal} {\bibinfo  {journal}
  {Angewandte Chemie International Edition}\ }\textbf {\bibinfo {volume}
  {42}},\ \bibinfo {pages} {5548} (\bibinfo {year} {2003})}\BibitemShut
  {NoStop}%
\bibitem [{\citenamefont {George}(2010)}]{george2010atomic}%
  \BibitemOpen
  \bibfield  {author} {\bibinfo {author} {\bibfnamefont {S.~M.}\ \bibnamefont
  {George}},\ }\href@noop {} {\bibfield  {journal} {\bibinfo  {journal} {Chem.
  Rev}\ }\textbf {\bibinfo {volume} {110}},\ \bibinfo {pages} {111} (\bibinfo
  {year} {2010})}\BibitemShut {NoStop}%
\bibitem [{\citenamefont {Stier}\ and\ \citenamefont
  {Nolting}(2011)}]{stier2011carrier}%
  \BibitemOpen
  \bibfield  {author} {\bibinfo {author} {\bibfnamefont {M.}~\bibnamefont
  {Stier}}\ and\ \bibinfo {author} {\bibfnamefont {W.}~\bibnamefont
  {Nolting}},\ }\href {\doibase 10.1103/PhysRevB.84.094417} {\bibfield
  {journal} {\bibinfo  {journal} {Phys. Rev. B}\ }\textbf {\bibinfo {volume}
  {84}},\ \bibinfo {pages} {094417} (\bibinfo {year} {2011})}\BibitemShut
  {NoStop}%
\bibitem [{\citenamefont {Smits}\ \emph {et~al.}(2004)\citenamefont {Smits},
  \citenamefont {Filip}, \citenamefont {Swagten}, \citenamefont {Koopmans},
  \citenamefont {De~Jonge}, \citenamefont {Chernyshova}, \citenamefont
  {Kowalczyk}, \citenamefont {Grasza}, \citenamefont {Szczerbakow},
  \citenamefont {Story} \emph {et~al.}}]{smits2004antiferromagnetic}%
  \BibitemOpen
  \bibfield  {author} {\bibinfo {author} {\bibfnamefont {C.}~\bibnamefont
  {Smits}}, \bibinfo {author} {\bibfnamefont {A.}~\bibnamefont {Filip}},
  \bibinfo {author} {\bibfnamefont {H.}~\bibnamefont {Swagten}}, \bibinfo
  {author} {\bibfnamefont {B.}~\bibnamefont {Koopmans}}, \bibinfo {author}
  {\bibfnamefont {W.}~\bibnamefont {De~Jonge}}, \bibinfo {author}
  {\bibfnamefont {M.}~\bibnamefont {Chernyshova}}, \bibinfo {author}
  {\bibfnamefont {L.}~\bibnamefont {Kowalczyk}}, \bibinfo {author}
  {\bibfnamefont {K.}~\bibnamefont {Grasza}}, \bibinfo {author} {\bibfnamefont
  {A.}~\bibnamefont {Szczerbakow}}, \bibinfo {author} {\bibfnamefont
  {T.}~\bibnamefont {Story}},  \emph {et~al.},\ }\href@noop {} {\bibfield
  {journal} {\bibinfo  {journal} {Physical Review B}\ }\textbf {\bibinfo
  {volume} {69}},\ \bibinfo {pages} {224410} (\bibinfo {year}
  {2004})}\BibitemShut {NoStop}%
\bibitem [{\citenamefont {Leiner}\ \emph {et~al.}(2010)\citenamefont {Leiner},
  \citenamefont {Lee}, \citenamefont {Yoo}, \citenamefont {Lee}, \citenamefont
  {Kirby}, \citenamefont {Tivakornsasithorn}, \citenamefont {Liu},
  \citenamefont {Furdyna},\ and\ \citenamefont
  {Dobrowolska}}]{leiner2010observation}%
  \BibitemOpen
  \bibfield  {author} {\bibinfo {author} {\bibfnamefont {J.}~\bibnamefont
  {Leiner}}, \bibinfo {author} {\bibfnamefont {H.}~\bibnamefont {Lee}},
  \bibinfo {author} {\bibfnamefont {T.}~\bibnamefont {Yoo}}, \bibinfo {author}
  {\bibfnamefont {S.}~\bibnamefont {Lee}}, \bibinfo {author} {\bibfnamefont
  {B.}~\bibnamefont {Kirby}}, \bibinfo {author} {\bibfnamefont
  {K.}~\bibnamefont {Tivakornsasithorn}}, \bibinfo {author} {\bibfnamefont
  {X.}~\bibnamefont {Liu}}, \bibinfo {author} {\bibfnamefont {J.}~\bibnamefont
  {Furdyna}}, \ and\ \bibinfo {author} {\bibfnamefont {M.}~\bibnamefont
  {Dobrowolska}},\ }\href@noop {} {\bibfield  {journal} {\bibinfo  {journal}
  {Physical Review B}\ }\textbf {\bibinfo {volume} {82}},\ \bibinfo {pages}
  {195205} (\bibinfo {year} {2010})}\BibitemShut {NoStop}%
\bibitem [{\citenamefont {Kepa}\ \emph {et~al.}(2001)\citenamefont {Kepa},
  \citenamefont {Kutner-Pielaszek}, \citenamefont {Blinowski}, \citenamefont
  {Twardowski}, \citenamefont {Majkrzak}, \citenamefont {Story}, \citenamefont
  {Kacman}, \citenamefont {Gałazka}, \citenamefont {Ha}, \citenamefont
  {Swagten}, \citenamefont {de~Jonge}, \citenamefont {Sipatov}, \citenamefont
  {Volobuev},\ and\ \citenamefont {Giebultowicz}}]{kepa2001antiferromagnetic}%
  \BibitemOpen
  \bibfield  {author} {\bibinfo {author} {\bibfnamefont {H.}~\bibnamefont
  {Kepa}}, \bibinfo {author} {\bibfnamefont {J.}~\bibnamefont
  {Kutner-Pielaszek}}, \bibinfo {author} {\bibfnamefont {J.}~\bibnamefont
  {Blinowski}}, \bibinfo {author} {\bibfnamefont {A.}~\bibnamefont
  {Twardowski}}, \bibinfo {author} {\bibfnamefont {C.~F.}\ \bibnamefont
  {Majkrzak}}, \bibinfo {author} {\bibfnamefont {T.}~\bibnamefont {Story}},
  \bibinfo {author} {\bibfnamefont {P.}~\bibnamefont {Kacman}}, \bibinfo
  {author} {\bibfnamefont {R.~R.}\ \bibnamefont {Gałazka}}, \bibinfo {author}
  {\bibfnamefont {K.}~\bibnamefont {Ha}}, \bibinfo {author} {\bibfnamefont
  {H.~J.~M.}\ \bibnamefont {Swagten}}, \bibinfo {author} {\bibfnamefont
  {W.~J.~M.}\ \bibnamefont {de~Jonge}}, \bibinfo {author} {\bibfnamefont
  {A.~Y.}\ \bibnamefont {Sipatov}}, \bibinfo {author} {\bibfnamefont
  {V.}~\bibnamefont {Volobuev}}, \ and\ \bibinfo {author} {\bibfnamefont
  {T.~M.}\ \bibnamefont {Giebultowicz}},\ }\href
  {http://stacks.iop.org/0295-5075/56/i=1/a=054} {\bibfield  {journal}
  {\bibinfo  {journal} {EPL (Europhysics Letters)}\ }\textbf {\bibinfo {volume}
  {56}},\ \bibinfo {pages} {54} (\bibinfo {year} {2001})}\BibitemShut {NoStop}%
\bibitem [{\citenamefont {Chern}\ \emph {et~al.}(2001)\citenamefont {Chern},
  \citenamefont {Horng}, \citenamefont {Shieh},\ and\ \citenamefont
  {Wu}}]{chern2001antiparallel}%
  \BibitemOpen
  \bibfield  {author} {\bibinfo {author} {\bibfnamefont {G.}~\bibnamefont
  {Chern}}, \bibinfo {author} {\bibfnamefont {L.}~\bibnamefont {Horng}},
  \bibinfo {author} {\bibfnamefont {W.~K.}\ \bibnamefont {Shieh}}, \ and\
  \bibinfo {author} {\bibfnamefont {T.~C.}\ \bibnamefont {Wu}},\ }\href
  {\doibase 10.1103/PhysRevB.63.094421} {\bibfield  {journal} {\bibinfo
  {journal} {Phys. Rev. B}\ }\textbf {\bibinfo {volume} {63}},\ \bibinfo
  {pages} {094421} (\bibinfo {year} {2001})}\BibitemShut {NoStop}%
\bibitem [{\citenamefont {Sankowski}\ and\ \citenamefont
  {Kacman}(2005)}]{sankowski2005interlayer}%
  \BibitemOpen
  \bibfield  {author} {\bibinfo {author} {\bibfnamefont {P.}~\bibnamefont
  {Sankowski}}\ and\ \bibinfo {author} {\bibfnamefont {P.}~\bibnamefont
  {Kacman}},\ }\href {\doibase 10.1103/PhysRevB.71.201303} {\bibfield
  {journal} {\bibinfo  {journal} {Phys. Rev. B}\ }\textbf {\bibinfo {volume}
  {71}},\ \bibinfo {pages} {201303} (\bibinfo {year} {2005})}\BibitemShut
  {NoStop}%
\bibitem [{\citenamefont {Chung}\ \emph {et~al.}(2011)\citenamefont {Chung},
  \citenamefont {Song}, \citenamefont {Yoo}, \citenamefont {Chung},
  \citenamefont {Lee}, \citenamefont {Kirby}, \citenamefont {Liu},\ and\
  \citenamefont {Furdyna}}]{chung2011investigation}%
  \BibitemOpen
  \bibfield  {author} {\bibinfo {author} {\bibfnamefont {J.-H.}\ \bibnamefont
  {Chung}}, \bibinfo {author} {\bibfnamefont {Y.-S.}\ \bibnamefont {Song}},
  \bibinfo {author} {\bibfnamefont {T.}~\bibnamefont {Yoo}}, \bibinfo {author}
  {\bibfnamefont {S.~J.}\ \bibnamefont {Chung}}, \bibinfo {author}
  {\bibfnamefont {S.}~\bibnamefont {Lee}}, \bibinfo {author} {\bibfnamefont
  {B.}~\bibnamefont {Kirby}}, \bibinfo {author} {\bibfnamefont
  {X.}~\bibnamefont {Liu}}, \ and\ \bibinfo {author} {\bibfnamefont
  {J.}~\bibnamefont {Furdyna}},\ }\href@noop {} {\bibfield  {journal} {\bibinfo
   {journal} {Journal of Applied Physics}\ }\textbf {\bibinfo {volume} {110}},\
  \bibinfo {pages} {013912} (\bibinfo {year} {2011})}\BibitemShut {NoStop}%
\bibitem [{\citenamefont {Samburskaya}\ \emph {et~al.}(2013)\citenamefont
  {Samburskaya}, \citenamefont {Sipatov}, \citenamefont {Volobuev},
  \citenamefont {Dziawa}, \citenamefont {Knoff}, \citenamefont {Kowalczyk},
  \citenamefont {Szot},\ and\ \citenamefont
  {Story}}]{samburskaya2013magnetization}%
  \BibitemOpen
  \bibfield  {author} {\bibinfo {author} {\bibfnamefont {T.}~\bibnamefont
  {Samburskaya}}, \bibinfo {author} {\bibfnamefont {A.~Y.}\ \bibnamefont
  {Sipatov}}, \bibinfo {author} {\bibfnamefont {V.}~\bibnamefont {Volobuev}},
  \bibinfo {author} {\bibfnamefont {P.}~\bibnamefont {Dziawa}}, \bibinfo
  {author} {\bibfnamefont {W.}~\bibnamefont {Knoff}}, \bibinfo {author}
  {\bibfnamefont {L.}~\bibnamefont {Kowalczyk}}, \bibinfo {author}
  {\bibfnamefont {M.}~\bibnamefont {Szot}}, \ and\ \bibinfo {author}
  {\bibfnamefont {T.}~\bibnamefont {Story}},\ }\href@noop {} {\bibfield
  {journal} {\bibinfo  {journal} {Acta Physica Polonica A}\ }\textbf {\bibinfo
  {volume} {124}},\ \bibinfo {pages} {133} (\bibinfo {year}
  {2013})}\BibitemShut {NoStop}%
\bibitem [{\citenamefont {Baxter}(1982)}]{baxter1982exactly}%
  \BibitemOpen
  \bibfield  {author} {\bibinfo {author} {\bibfnamefont {R.}~\bibnamefont
  {Baxter}},\ }\href@noop {} {\emph {\bibinfo {title} {Exactly Solved Models in
  Statistical Mechanics}}},\ Vol.~\bibinfo {volume} {9}\ (\bibinfo  {publisher}
  {Academic press London},\ \bibinfo {address} {London, UK},\ \bibinfo {year}
  {1982})\BibitemShut {NoStop}%
\bibitem [{\citenamefont {Lipowski}\ and\ \citenamefont
  {Suzuki}(1993)}]{lipowski1993layered}%
  \BibitemOpen
  \bibfield  {author} {\bibinfo {author} {\bibfnamefont {A.}~\bibnamefont
  {Lipowski}}\ and\ \bibinfo {author} {\bibfnamefont {M.}~\bibnamefont
  {Suzuki}},\ }\href@noop {} {\bibfield  {journal} {\bibinfo  {journal}
  {Physica A: Statistical Mechanics and its Applications}\ }\textbf {\bibinfo
  {volume} {198}},\ \bibinfo {pages} {227} (\bibinfo {year}
  {1993})}\BibitemShut {NoStop}%
\bibitem [{\citenamefont {Hansen}\ \emph {et~al.}(1993)\citenamefont {Hansen},
  \citenamefont {Lemmich}, \citenamefont {Ipsen},\ and\ \citenamefont
  {Mouritsen}}]{hansen1993two}%
  \BibitemOpen
  \bibfield  {author} {\bibinfo {author} {\bibfnamefont {P.~L.}\ \bibnamefont
  {Hansen}}, \bibinfo {author} {\bibfnamefont {J.}~\bibnamefont {Lemmich}},
  \bibinfo {author} {\bibfnamefont {J.~H.}\ \bibnamefont {Ipsen}}, \ and\
  \bibinfo {author} {\bibfnamefont {O.~G.}\ \bibnamefont {Mouritsen}},\
  }\href@noop {} {\bibfield  {journal} {\bibinfo  {journal} {Journal of
  Statistical Physics}\ }\textbf {\bibinfo {volume} {73}},\ \bibinfo {pages}
  {723} (\bibinfo {year} {1993})}\BibitemShut {NoStop}%
\bibitem [{\citenamefont {Kaneyoshi}(1995)}]{kaneyoshi1995relation}%
  \BibitemOpen
  \bibfield  {author} {\bibinfo {author} {\bibfnamefont {T.}~\bibnamefont
  {Kaneyoshi}},\ }\href {\doibase https://doi.org/10.1016/0038-1098(94)00752-7}
  {\bibfield  {journal} {\bibinfo  {journal} {Solid State Communications}\
  }\textbf {\bibinfo {volume} {93}},\ \bibinfo {pages} {691 } (\bibinfo {year}
  {1995})}\BibitemShut {NoStop}%
\bibitem [{\citenamefont {Oitmaa}(2005)}]{oitimaa2005ferrimagnetism}%
  \BibitemOpen
  \bibfield  {author} {\bibinfo {author} {\bibfnamefont {J.}~\bibnamefont
  {Oitmaa}},\ }\href {\doibase 10.1103/PhysRevB.72.224404} {\bibfield
  {journal} {\bibinfo  {journal} {Phys. Rev. B}\ }\textbf {\bibinfo {volume}
  {72}},\ \bibinfo {pages} {224404} (\bibinfo {year} {2005})}\BibitemShut
  {NoStop}%
\bibitem [{\citenamefont {Kaneyoshi}\ and\ \citenamefont
  {Ja\v{s}\v{c}ur}(1993)}]{kaneyoshi1993magnetic}%
  \BibitemOpen
  \bibfield  {author} {\bibinfo {author} {\bibfnamefont {T.}~\bibnamefont
  {Kaneyoshi}}\ and\ \bibinfo {author} {\bibfnamefont {M.}~\bibnamefont
  {Ja\v{s}\v{c}ur}},\ }\href {\doibase
  https://doi.org/10.1016/0378-4371(93)90171-Y} {\bibfield  {journal} {\bibinfo
   {journal} {Physica A: Statistical Mechanics and its Applications}\ }\textbf
  {\bibinfo {volume} {195}},\ \bibinfo {pages} {474 } (\bibinfo {year}
  {1993})}\BibitemShut {NoStop}%
\bibitem [{\citenamefont {Ja{\v{s}}{\v{c}}ur}\ and\ \citenamefont
  {Kaneyoshi}(1995)}]{javsvcur1995effect}%
  \BibitemOpen
  \bibfield  {author} {\bibinfo {author} {\bibfnamefont {M.}~\bibnamefont
  {Ja{\v{s}}{\v{c}}ur}}\ and\ \bibinfo {author} {\bibfnamefont
  {T.}~\bibnamefont {Kaneyoshi}},\ }\href@noop {} {\bibfield  {journal}
  {\bibinfo  {journal} {Physica A: Statistical Mechanics and its Applications}\
  }\textbf {\bibinfo {volume} {220}},\ \bibinfo {pages} {542} (\bibinfo {year}
  {1995})}\BibitemShut {NoStop}%
\bibitem [{\citenamefont {Ainane}\ \emph {et~al.}(2007)\citenamefont {Ainane},
  \citenamefont {H{\"a}ussler}, \citenamefont {Htoutou},\ and\ \citenamefont
  {Saber}}]{ainane2007magnetic}%
  \BibitemOpen
  \bibfield  {author} {\bibinfo {author} {\bibfnamefont {A.}~\bibnamefont
  {Ainane}}, \bibinfo {author} {\bibfnamefont {P.}~\bibnamefont
  {H{\"a}ussler}}, \bibinfo {author} {\bibfnamefont {K.}~\bibnamefont
  {Htoutou}}, \ and\ \bibinfo {author} {\bibfnamefont {M.}~\bibnamefont
  {Saber}},\ }\href@noop {} {\bibfield  {journal} {\bibinfo  {journal} {Surface
  Science}\ }\textbf {\bibinfo {volume} {601}},\ \bibinfo {pages} {4256}
  (\bibinfo {year} {2007})}\BibitemShut {NoStop}%
\bibitem [{\citenamefont {Deviren}\ \emph {et~al.}(2008)\citenamefont
  {Deviren}, \citenamefont {Canko},\ and\ \citenamefont
  {Keskin}}]{bayram2008effective}%
  \BibitemOpen
  \bibfield  {author} {\bibinfo {author} {\bibfnamefont {B.}~\bibnamefont
  {Deviren}}, \bibinfo {author} {\bibfnamefont {O.}~\bibnamefont {Canko}}, \
  and\ \bibinfo {author} {\bibfnamefont {M.}~\bibnamefont {Keskin}},\ }\href
  {\doibase https://doi.org/10.1016/j.jmmm.2008.04.130} {\bibfield  {journal}
  {\bibinfo  {journal} {Journal of Magnetism and Magnetic Materials}\ }\textbf
  {\bibinfo {volume} {320}},\ \bibinfo {pages} {2291 } (\bibinfo {year}
  {2008})}\BibitemShut {NoStop}%
\bibitem [{\citenamefont {Deviren}\ \emph
  {et~al.}(2011{\natexlab{a}})\citenamefont {Deviren}, \citenamefont
  {Akbudak},\ and\ \citenamefont {Keskin}}]{bayram2011mixed}%
  \BibitemOpen
  \bibfield  {author} {\bibinfo {author} {\bibfnamefont {B.}~\bibnamefont
  {Deviren}}, \bibinfo {author} {\bibfnamefont {S.}~\bibnamefont {Akbudak}}, \
  and\ \bibinfo {author} {\bibfnamefont {M.}~\bibnamefont {Keskin}},\ }\href
  {\doibase https://doi.org/10.1016/j.ssc.2010.11.039} {\bibfield  {journal}
  {\bibinfo  {journal} {Solid State Communications}\ }\textbf {\bibinfo
  {volume} {151}},\ \bibinfo {pages} {193 } (\bibinfo {year}
  {2011}{\natexlab{a}})}\BibitemShut {NoStop}%
\bibitem [{\citenamefont {Deviren}\ \emph
  {et~al.}(2011{\natexlab{b}})\citenamefont {Deviren}, \citenamefont {Polat},\
  and\ \citenamefont {Keskin}}]{bayram2011phase}%
  \BibitemOpen
  \bibfield  {author} {\bibinfo {author} {\bibfnamefont {B.}~\bibnamefont
  {Deviren}}, \bibinfo {author} {\bibfnamefont {Y.}~\bibnamefont {Polat}}, \
  and\ \bibinfo {author} {\bibfnamefont {M.}~\bibnamefont {Keskin}},\ }\href
  {http://stacks.iop.org/1674-1056/20/i=6/a=060507} {\bibfield  {journal}
  {\bibinfo  {journal} {Chinese Physics B}\ }\textbf {\bibinfo {volume} {20}},\
  \bibinfo {pages} {060507} (\bibinfo {year} {2011}{\natexlab{b}})}\BibitemShut
  {NoStop}%
\bibitem [{\citenamefont {Kantar}\ and\ \citenamefont
  {Ertaş}(2014)}]{ersin2014magnetic}%
  \BibitemOpen
  \bibfield  {author} {\bibinfo {author} {\bibfnamefont {E.}~\bibnamefont
  {Kantar}}\ and\ \bibinfo {author} {\bibfnamefont {M.}~\bibnamefont
  {Ertaş}},\ }\href {\doibase https://doi.org/10.1016/j.ssc.2014.03.006}
  {\bibfield  {journal} {\bibinfo  {journal} {Solid State Communications}\
  }\textbf {\bibinfo {volume} {188}},\ \bibinfo {pages} {71 } (\bibinfo {year}
  {2014})}\BibitemShut {NoStop}%
\bibitem [{\citenamefont {Li}\ \emph {et~al.}(2001)\citenamefont {Li},
  \citenamefont {Shuai}, \citenamefont {Wang}, \citenamefont {Luo},\ and\
  \citenamefont {Sch{\"u}lke}}]{li2001critical}%
  \BibitemOpen
  \bibfield  {author} {\bibinfo {author} {\bibfnamefont {Z.}~\bibnamefont
  {Li}}, \bibinfo {author} {\bibfnamefont {Z.}~\bibnamefont {Shuai}}, \bibinfo
  {author} {\bibfnamefont {Q.}~\bibnamefont {Wang}}, \bibinfo {author}
  {\bibfnamefont {H.}~\bibnamefont {Luo}}, \ and\ \bibinfo {author}
  {\bibfnamefont {L.}~\bibnamefont {Sch{\"u}lke}},\ }\href@noop {} {\bibfield
  {journal} {\bibinfo  {journal} {Journal of Physics A: Mathematical and
  General}\ }\textbf {\bibinfo {volume} {34}},\ \bibinfo {pages} {6069}
  (\bibinfo {year} {2001})}\BibitemShut {NoStop}%
\bibitem [{\citenamefont {Mirza}\ and\ \citenamefont
  {Mardani}(2003)}]{mirza2003phenomenological}%
  \BibitemOpen
  \bibfield  {author} {\bibinfo {author} {\bibfnamefont {B.}~\bibnamefont
  {Mirza}}\ and\ \bibinfo {author} {\bibfnamefont {T.}~\bibnamefont
  {Mardani}},\ }\href@noop {} {\bibfield  {journal} {\bibinfo  {journal} {The
  European Physical Journal B-Condensed Matter and Complex Systems}\ }\textbf
  {\bibinfo {volume} {34}},\ \bibinfo {pages} {321} (\bibinfo {year}
  {2003})}\BibitemShut {NoStop}%
\bibitem [{\citenamefont {Lipowski}(1998)}]{lipowski1998critical}%
  \BibitemOpen
  \bibfield  {author} {\bibinfo {author} {\bibfnamefont {A.}~\bibnamefont
  {Lipowski}},\ }\href@noop {} {\bibfield  {journal} {\bibinfo  {journal}
  {Physica A: Statistical Mechanics and its Applications}\ }\textbf {\bibinfo
  {volume} {250}},\ \bibinfo {pages} {373} (\bibinfo {year}
  {1998})}\BibitemShut {NoStop}%
\bibitem [{\citenamefont {Ferrenberg}\ and\ \citenamefont
  {Landau}(1991{\natexlab{a}})}]{ferrenberg1991monte}%
  \BibitemOpen
  \bibfield  {author} {\bibinfo {author} {\bibfnamefont {A.~M.}\ \bibnamefont
  {Ferrenberg}}\ and\ \bibinfo {author} {\bibfnamefont {D.}~\bibnamefont
  {Landau}},\ }\href@noop {} {\bibfield  {journal} {\bibinfo  {journal}
  {Journal of applied physics}\ }\textbf {\bibinfo {volume} {70}},\ \bibinfo
  {pages} {6215} (\bibinfo {year} {1991}{\natexlab{a}})}\BibitemShut {NoStop}%
\bibitem [{\citenamefont {Zaim}\ \emph {et~al.}(2013)\citenamefont {Zaim},
  \citenamefont {Kerouad},\ and\ \citenamefont {Boughrara}}]{ahmed2013monte}%
  \BibitemOpen
  \bibfield  {author} {\bibinfo {author} {\bibfnamefont {A.}~\bibnamefont
  {Zaim}}, \bibinfo {author} {\bibfnamefont {M.}~\bibnamefont {Kerouad}}, \
  and\ \bibinfo {author} {\bibfnamefont {M.}~\bibnamefont {Boughrara}},\ }\href
  {\doibase https://doi.org/10.1016/j.ssc.2012.10.014} {\bibfield  {journal}
  {\bibinfo  {journal} {Solid State Communications}\ }\textbf {\bibinfo
  {volume} {158}},\ \bibinfo {pages} {76 } (\bibinfo {year}
  {2013})}\BibitemShut {NoStop}%
\bibitem [{\citenamefont {Wang}\ \emph {et~al.}(2016)\citenamefont {Wang},
  \citenamefont {Liu}, \citenamefont {Lv},\ and\ \citenamefont
  {Luo}}]{wang2016monte}%
  \BibitemOpen
  \bibfield  {author} {\bibinfo {author} {\bibfnamefont {W.}~\bibnamefont
  {Wang}}, \bibinfo {author} {\bibfnamefont {R.}~\bibnamefont {Liu}}, \bibinfo
  {author} {\bibfnamefont {D.}~\bibnamefont {Lv}}, \ and\ \bibinfo {author}
  {\bibfnamefont {X.}~\bibnamefont {Luo}},\ }\href {\doibase
  http://dx.doi.org/10.1016/j.spmi.2016.08.045} {\bibfield  {journal} {\bibinfo
   {journal} {Superlattices and Microstructures}\ }\textbf {\bibinfo {volume}
  {98}},\ \bibinfo {pages} {458} (\bibinfo {year} {2016})}\BibitemShut
  {NoStop}%
\bibitem [{\citenamefont {Wang}\ \emph {et~al.}(2017)\citenamefont {Wang},
  \citenamefont {Xue},\ and\ \citenamefont {Wang}}]{wang2017compensation}%
  \BibitemOpen
  \bibfield  {author} {\bibinfo {author} {\bibfnamefont {W.}~\bibnamefont
  {Wang}}, \bibinfo {author} {\bibfnamefont {F.-l.}\ \bibnamefont {Xue}}, \
  and\ \bibinfo {author} {\bibfnamefont {M.-z.}\ \bibnamefont {Wang}},\ }\href
  {\doibase http://dx.doi.org/10.1016/j.physb.2017.04.001} {\bibfield
  {journal} {\bibinfo  {journal} {Physica B: Condensed Matter}\ }\textbf
  {\bibinfo {volume} {515}},\ \bibinfo {pages} {104} (\bibinfo {year}
  {2017})}\BibitemShut {NoStop}%
\bibitem [{\citenamefont {Diaz}\ and\ \citenamefont
  {Branco}(2017)}]{diaz2017monte}%
  \BibitemOpen
  \bibfield  {author} {\bibinfo {author} {\bibfnamefont {I.~J.~L.}\
  \bibnamefont {Diaz}}\ and\ \bibinfo {author} {\bibfnamefont {N.~S.}\
  \bibnamefont {Branco}},\ }\href {\doibase
  https://doi.org/10.1016/j.physa.2016.10.055} {\bibfield  {journal} {\bibinfo
  {journal} {Physica A: Statistical Mechanics and its Applications}\ }\textbf
  {\bibinfo {volume} {468}},\ \bibinfo {pages} {158} (\bibinfo {year}
  {2017})}\BibitemShut {NoStop}%
\bibitem [{\citenamefont {Asgari}\ and\ \citenamefont
  {Ghaemi}(2008)}]{asgari2008obtaining}%
  \BibitemOpen
  \bibfield  {author} {\bibinfo {author} {\bibfnamefont {Y.}~\bibnamefont
  {Asgari}}\ and\ \bibinfo {author} {\bibfnamefont {M.}~\bibnamefont
  {Ghaemi}},\ }\href@noop {} {\bibfield  {journal} {\bibinfo  {journal}
  {Physica A: Statistical Mechanics and its Applications}\ }\textbf {\bibinfo
  {volume} {387}},\ \bibinfo {pages} {1937} (\bibinfo {year}
  {2008})}\BibitemShut {NoStop}%
\bibitem [{\citenamefont {Xu}\ and\ \citenamefont
  {Du}(2017)}]{ping2017magnetization}%
  \BibitemOpen
  \bibfield  {author} {\bibinfo {author} {\bibfnamefont {P.}~\bibnamefont
  {Xu}}\ and\ \bibinfo {author} {\bibfnamefont {A.}~\bibnamefont {Du}},\ }\href
  {\doibase https://doi.org/10.1016/j.physb.2017.06.052} {\bibfield  {journal}
  {\bibinfo  {journal} {Physica B: Condensed Matter}\ }\textbf {\bibinfo
  {volume} {521}},\ \bibinfo {pages} {134 } (\bibinfo {year}
  {2017})}\BibitemShut {NoStop}%
\bibitem [{\citenamefont {Sza{\l}owski}\ and\ \citenamefont
  {Balcerzak}(2013)}]{szalowski2013influence}%
  \BibitemOpen
  \bibfield  {author} {\bibinfo {author} {\bibfnamefont {K.}~\bibnamefont
  {Sza{\l}owski}}\ and\ \bibinfo {author} {\bibfnamefont {T.}~\bibnamefont
  {Balcerzak}},\ }\href@noop {} {\bibfield  {journal} {\bibinfo  {journal}
  {Thin Solid Films}\ }\textbf {\bibinfo {volume} {534}},\ \bibinfo {pages}
  {546} (\bibinfo {year} {2013})}\BibitemShut {NoStop}%
\bibitem [{\citenamefont {Balcerzak}\ and\ \citenamefont
  {Sza{\l}owski}(2014)}]{balcerzak2014ferrimagnetism}%
  \BibitemOpen
  \bibfield  {author} {\bibinfo {author} {\bibfnamefont {T.}~\bibnamefont
  {Balcerzak}}\ and\ \bibinfo {author} {\bibfnamefont {K.}~\bibnamefont
  {Sza{\l}owski}},\ }\href@noop {} {\bibfield  {journal} {\bibinfo  {journal}
  {Physica A: Statistical Mechanics and its Applications}\ }\textbf {\bibinfo
  {volume} {395}},\ \bibinfo {pages} {183} (\bibinfo {year}
  {2014})}\BibitemShut {NoStop}%
\bibitem [{\citenamefont {Sza{\l}owski}\ and\ \citenamefont
  {Balcerzak}(2012)}]{szalowski2012critical}%
  \BibitemOpen
  \bibfield  {author} {\bibinfo {author} {\bibfnamefont {K.}~\bibnamefont
  {Sza{\l}owski}}\ and\ \bibinfo {author} {\bibfnamefont {T.}~\bibnamefont
  {Balcerzak}},\ }\href@noop {} {\bibfield  {journal} {\bibinfo  {journal}
  {Physica A: Statistical Mechanics and its Applications}\ }\textbf {\bibinfo
  {volume} {391}},\ \bibinfo {pages} {2197} (\bibinfo {year}
  {2012})}\BibitemShut {NoStop}%
\bibitem [{\citenamefont {Sza{\l}owski}\ and\ \citenamefont
  {Balcerzak}(2014)}]{szalowski2014normal}%
  \BibitemOpen
  \bibfield  {author} {\bibinfo {author} {\bibfnamefont {K.}~\bibnamefont
  {Sza{\l}owski}}\ and\ \bibinfo {author} {\bibfnamefont {T.}~\bibnamefont
  {Balcerzak}},\ }\href@noop {} {\bibfield  {journal} {\bibinfo  {journal}
  {Journal of Physics: Condensed Matter}\ }\textbf {\bibinfo {volume} {26}},\
  \bibinfo {pages} {386003} (\bibinfo {year} {2014})}\BibitemShut {NoStop}%
\bibitem [{\citenamefont {Razouk}\ \emph {et~al.}(2011)\citenamefont {Razouk},
  \citenamefont {Sahlaoui},\ and\ \citenamefont
  {Sajieddine}}]{razouk2011monte}%
  \BibitemOpen
  \bibfield  {author} {\bibinfo {author} {\bibfnamefont {A.}~\bibnamefont
  {Razouk}}, \bibinfo {author} {\bibfnamefont {M.}~\bibnamefont {Sahlaoui}}, \
  and\ \bibinfo {author} {\bibfnamefont {M.}~\bibnamefont {Sajieddine}},\
  }\href@noop {} {\bibfield  {journal} {\bibinfo  {journal} {Journal of
  superconductivity and novel magnetism}\ }\textbf {\bibinfo {volume} {24}},\
  \bibinfo {pages} {1901} (\bibinfo {year} {2011})}\BibitemShut {NoStop}%
\bibitem [{\citenamefont {Diaz}\ and\ \citenamefont
  {Branco}(2018{\natexlab{a}})}]{diaz2018monte}%
  \BibitemOpen
  \bibfield  {author} {\bibinfo {author} {\bibfnamefont {I.~J.~L.}\
  \bibnamefont {Diaz}}\ and\ \bibinfo {author} {\bibfnamefont {N.~S.}\
  \bibnamefont {Branco}},\ }\href {\doibase
  https://doi.org/10.1016/j.physa.2017.09.005} {\bibfield  {journal} {\bibinfo
  {journal} {{Physica A: Statistical Mechanics and its Applications}}\ }\textbf
  {\bibinfo {volume} {490}},\ \bibinfo {pages} {904} (\bibinfo {year}
  {2018}{\natexlab{a}})}\BibitemShut {NoStop}%
\bibitem [{\citenamefont {Naji}\ \emph
  {et~al.}(2014{\natexlab{a}})\citenamefont {Naji}, \citenamefont {Belhaj},
  \citenamefont {Labrim}, \citenamefont {Bahmad}, \citenamefont {Benyoussef},\
  and\ \citenamefont {El~Kenz}}]{naji2014phase}%
  \BibitemOpen
  \bibfield  {author} {\bibinfo {author} {\bibfnamefont {S.}~\bibnamefont
  {Naji}}, \bibinfo {author} {\bibfnamefont {A.}~\bibnamefont {Belhaj}},
  \bibinfo {author} {\bibfnamefont {H.}~\bibnamefont {Labrim}}, \bibinfo
  {author} {\bibfnamefont {L.}~\bibnamefont {Bahmad}}, \bibinfo {author}
  {\bibfnamefont {A.}~\bibnamefont {Benyoussef}}, \ and\ \bibinfo {author}
  {\bibfnamefont {A.}~\bibnamefont {El~Kenz}},\ }\href@noop {} {\bibfield
  {journal} {\bibinfo  {journal} {Physica A: Statistical Mechanics and its
  Applications}\ }\textbf {\bibinfo {volume} {399}},\ \bibinfo {pages} {106}
  (\bibinfo {year} {2014}{\natexlab{a}})}\BibitemShut {NoStop}%
\bibitem [{\citenamefont {Naji}\ \emph
  {et~al.}(2014{\natexlab{b}})\citenamefont {Naji}, \citenamefont {Belhaj},
  \citenamefont {Labrim}, \citenamefont {Bahmad}, \citenamefont {Benyoussef},\
  and\ \citenamefont {El~Kenz}}]{naji2014monte}%
  \BibitemOpen
  \bibfield  {author} {\bibinfo {author} {\bibfnamefont {S.}~\bibnamefont
  {Naji}}, \bibinfo {author} {\bibfnamefont {A.}~\bibnamefont {Belhaj}},
  \bibinfo {author} {\bibfnamefont {H.}~\bibnamefont {Labrim}}, \bibinfo
  {author} {\bibfnamefont {L.}~\bibnamefont {Bahmad}}, \bibinfo {author}
  {\bibfnamefont {A.}~\bibnamefont {Benyoussef}}, \ and\ \bibinfo {author}
  {\bibfnamefont {A.}~\bibnamefont {El~Kenz}},\ }\href@noop {} {\bibfield
  {journal} {\bibinfo  {journal} {Acta Phys Pol Ser B}\ }\textbf {\bibinfo
  {volume} {45}},\ \bibinfo {pages} {947} (\bibinfo {year}
  {2014}{\natexlab{b}})}\BibitemShut {NoStop}%
\bibitem [{\citenamefont {Santos}\ and\ \citenamefont
  {Barreto}(2017)}]{santos2017effective}%
  \BibitemOpen
  \bibfield  {author} {\bibinfo {author} {\bibfnamefont {J.~P.}\ \bibnamefont
  {Santos}}\ and\ \bibinfo {author} {\bibfnamefont {F.~S.}\ \bibnamefont
  {Barreto}},\ }\href {\doibase https://doi.org/10.1016/j.jmmm.2017.05.017}
  {\bibfield  {journal} {\bibinfo  {journal} {{Journal of Magnetism and
  Magnetic Materials}}\ }\textbf {\bibinfo {volume} {439}},\ \bibinfo {pages}
  {114 } (\bibinfo {year} {2017})}\BibitemShut {NoStop}%
\bibitem [{\citenamefont {Diaz}\ and\ \citenamefont
  {Branco}(2018{\natexlab{b}})}]{diaz2018ferrimagnetism}%
  \BibitemOpen
  \bibfield  {author} {\bibinfo {author} {\bibfnamefont {I.~J.~L.}\
  \bibnamefont {Diaz}}\ and\ \bibinfo {author} {\bibfnamefont {N.~S.}\
  \bibnamefont {Branco}},\ }\href {\doibase
  https://doi.org/10.1016/j.physb.2017.10.036} {\bibfield  {journal} {\bibinfo
  {journal} {Physica B: Condensed Matter}\ }\textbf {\bibinfo {volume} {529}},\
  \bibinfo {pages} {73 } (\bibinfo {year} {2018}{\natexlab{b}})},\ \Eprint
  {http://arxiv.org/abs/1710.10298} {arXiv:1710.10298 [cond-mat]} \BibitemShut
  {NoStop}%
\bibitem [{\citenamefont {Boechat}\ \emph {et~al.}(2002)\citenamefont
  {Boechat}, \citenamefont {Filgueiras}, \citenamefont {Cordeiro},\ and\
  \citenamefont {Branco}}]{boechat2002renormalization}%
  \BibitemOpen
  \bibfield  {author} {\bibinfo {author} {\bibfnamefont {B.}~\bibnamefont
  {Boechat}}, \bibinfo {author} {\bibfnamefont {R.}~\bibnamefont {Filgueiras}},
  \bibinfo {author} {\bibfnamefont {C.}~\bibnamefont {Cordeiro}}, \ and\
  \bibinfo {author} {\bibfnamefont {N.}~\bibnamefont {Branco}},\ }\href
  {\doibase http://dx.doi.org/10.1016/S0378-4371(01)00560-X} {\bibfield
  {journal} {\bibinfo  {journal} {Physica A: Statistical Mechanics and its
  Applications}\ }\textbf {\bibinfo {volume} {304}},\ \bibinfo {pages} {429 }
  (\bibinfo {year} {2002})}\BibitemShut {NoStop}%
\bibitem [{\citenamefont {Wolff}(1989)}]{artigo:wolff}%
  \BibitemOpen
  \bibfield  {author} {\bibinfo {author} {\bibfnamefont {U.}~\bibnamefont
  {Wolff}},\ }\href {\doibase 10.1103/PhysRevLett.62.361} {\bibfield  {journal}
  {\bibinfo  {journal} {Phys. Rev. Lett.}\ }\textbf {\bibinfo {volume} {62}},\
  \bibinfo {pages} {361} (\bibinfo {year} {1989})}\BibitemShut {NoStop}%
\bibitem [{\citenamefont {Ferrenberg}\ and\ \citenamefont
  {Swendsen}(1988)}]{artigo:ferrenberg:histograma1}%
  \BibitemOpen
  \bibfield  {author} {\bibinfo {author} {\bibfnamefont {A.~M.}\ \bibnamefont
  {Ferrenberg}}\ and\ \bibinfo {author} {\bibfnamefont {R.~H.}\ \bibnamefont
  {Swendsen}},\ }\href {\doibase 10.1103/PhysRevLett.61.2635} {\bibfield
  {journal} {\bibinfo  {journal} {Phys. Rev. Lett.}\ }\textbf {\bibinfo
  {volume} {61}},\ \bibinfo {pages} {2635} (\bibinfo {year}
  {1988})}\BibitemShut {NoStop}%
\bibitem [{\citenamefont {Ferrenberg}\ and\ \citenamefont
  {Swendsen}(1989)}]{artigo:ferrenberg:histograma2}%
  \BibitemOpen
  \bibfield  {author} {\bibinfo {author} {\bibfnamefont {A.~M.}\ \bibnamefont
  {Ferrenberg}}\ and\ \bibinfo {author} {\bibfnamefont {R.~H.}\ \bibnamefont
  {Swendsen}},\ }\href {\doibase 10.1103/PhysRevLett.63.1195} {\bibfield
  {journal} {\bibinfo  {journal} {Phys. Rev. Lett.}\ }\textbf {\bibinfo
  {volume} {63}},\ \bibinfo {pages} {1195} (\bibinfo {year}
  {1989})}\BibitemShut {NoStop}%
\bibitem [{\citenamefont {Metropolis}\ \emph {et~al.}(1953)\citenamefont
  {Metropolis}, \citenamefont {Rosenbluth}, \citenamefont {Rosenbluth},
  \citenamefont {Teller},\ and\ \citenamefont {Teller}}]{artigo:metropolis}%
  \BibitemOpen
  \bibfield  {author} {\bibinfo {author} {\bibfnamefont {N.}~\bibnamefont
  {Metropolis}}, \bibinfo {author} {\bibfnamefont {A.}~\bibnamefont
  {Rosenbluth}}, \bibinfo {author} {\bibfnamefont {M.}~\bibnamefont
  {Rosenbluth}}, \bibinfo {author} {\bibfnamefont {A.}~\bibnamefont {Teller}},
  \ and\ \bibinfo {author} {\bibfnamefont {E.}~\bibnamefont {Teller}},\
  }\href@noop {} {\bibfield  {journal} {\bibinfo  {journal} {J. Chem. Phys.}\
  }\textbf {\bibinfo {volume} {21}},\ \bibinfo {pages} {1087} (\bibinfo {year}
  {1953})}\BibitemShut {NoStop}%
\bibitem [{\citenamefont {Matsumoto}\ and\ \citenamefont
  {Nishimura}(1998)}]{artigo:mersenne-twister}%
  \BibitemOpen
  \bibfield  {author} {\bibinfo {author} {\bibfnamefont {M.}~\bibnamefont
  {Matsumoto}}\ and\ \bibinfo {author} {\bibfnamefont {T.}~\bibnamefont
  {Nishimura}},\ }\href@noop {} {\bibfield  {journal} {\bibinfo  {journal} {ACM
  Transactions on Modeling and Computer Simulation (TOMACS)}\ }\textbf
  {\bibinfo {volume} {8}},\ \bibinfo {pages} {3} (\bibinfo {year}
  {1998})}\BibitemShut {NoStop}%
\bibitem [{\citenamefont {Newman}\ and\ \citenamefont
  {Barkema}(1999)}]{livro:barkema}%
  \BibitemOpen
  \bibfield  {author} {\bibinfo {author} {\bibfnamefont {M.~E.~J.}\
  \bibnamefont {Newman}}\ and\ \bibinfo {author} {\bibfnamefont {G.~T.}\
  \bibnamefont {Barkema}},\ }\href@noop {} {\emph {\bibinfo {title}
  {\uppercase{M}onte \uppercase{C}arlo Methods in Statistical Physics}}}\
  (\bibinfo  {publisher} {Oxford University Press},\ \bibinfo {address} {New
  York, USA},\ \bibinfo {year} {1999})\BibitemShut {NoStop}%
\bibitem [{\citenamefont {Yeomans}(1992)}]{livro:julia}%
  \BibitemOpen
  \bibfield  {author} {\bibinfo {author} {\bibfnamefont {J.}~\bibnamefont
  {Yeomans}},\ }\href@noop {} {\emph {\bibinfo {title} {Statistcal Mechanics of
  Phase Transtions}}}\ (\bibinfo  {publisher} {Clarendon Press},\ \bibinfo
  {address} {New York, USA},\ \bibinfo {year} {1992})\BibitemShut {NoStop}%
\bibitem [{\citenamefont {Ferrenberg}\ and\ \citenamefont
  {Landau}(1991{\natexlab{b}})}]{artigo:landau}%
  \BibitemOpen
  \bibfield  {author} {\bibinfo {author} {\bibfnamefont {A.~M.}\ \bibnamefont
  {Ferrenberg}}\ and\ \bibinfo {author} {\bibfnamefont {D.~P.}\ \bibnamefont
  {Landau}},\ }\href {\doibase 10.1103/PhysRevB.44.5081} {\bibfield  {journal}
  {\bibinfo  {journal} {Phys. Rev. B}\ }\textbf {\bibinfo {volume} {44}},\
  \bibinfo {pages} {5081} (\bibinfo {year} {1991}{\natexlab{b}})}\BibitemShut
  {NoStop}%
\bibitem [{\citenamefont {Broyden}(1970{\natexlab{a}})}]{artigo:BFGS:1}%
  \BibitemOpen
  \bibfield  {author} {\bibinfo {author} {\bibfnamefont {C.~G.}\ \bibnamefont
  {Broyden}},\ }\href@noop {} {\bibfield  {journal} {\bibinfo  {journal} {IMA
  Journal of Applied Mathematics}\ }\textbf {\bibinfo {volume} {6}},\ \bibinfo
  {pages} {76} (\bibinfo {year} {1970}{\natexlab{a}})}\BibitemShut {NoStop}%
\bibitem [{\citenamefont {Broyden}(1970{\natexlab{b}})}]{artigo:BFGS:2}%
  \BibitemOpen
  \bibfield  {author} {\bibinfo {author} {\bibfnamefont {C.~G.}\ \bibnamefont
  {Broyden}},\ }\href@noop {} {\bibfield  {journal} {\bibinfo  {journal} {IMA
  journal of applied mathematics}\ }\textbf {\bibinfo {volume} {6}},\ \bibinfo
  {pages} {222} (\bibinfo {year} {1970}{\natexlab{b}})}\BibitemShut {NoStop}%
\bibitem [{\citenamefont {Diaz}\ and\ \citenamefont
  {Branco}(2012)}]{diaz2012feru}%
  \BibitemOpen
  \bibfield  {author} {\bibinfo {author} {\bibfnamefont {I.~J.~L.}\
  \bibnamefont {Diaz}}\ and\ \bibinfo {author} {\bibfnamefont {N.~S.}\
  \bibnamefont {Branco}},\ }\href {\doibase 10.1103/PhysRevE.85.021142}
  {\bibfield  {journal} {\bibinfo  {journal} {Physical Review E}\ }\textbf
  {\bibinfo {volume} {85}},\ \bibinfo {pages} {021142} (\bibinfo {year}
  {2012})}\BibitemShut {NoStop}%
\bibitem [{\citenamefont {Brent}(1973)}]{artigo:brent1973}%
  \BibitemOpen
  \bibfield  {author} {\bibinfo {author} {\bibfnamefont {R.~P.}\ \bibnamefont
  {Brent}},\ }\href@noop {} {\bibfield  {journal} {\bibinfo  {journal} {SIAM
  Journal on Numerical Analysis}\ }\textbf {\bibinfo {volume} {10}},\ \bibinfo
  {pages} {327} (\bibinfo {year} {1973})}\BibitemShut {NoStop}%
\end{thebibliography}
\end{document}